%% file: main.tex
\documentclass[journal,12]{IEEEtran}
\usepackage[pdftex]{graphicx}
\graphicspath{{../pdf/}{../jpeg/}}
\DeclareGraphicsExtensions{.pdf,.jpeg,.png}
\usepackage[cmex10]{amsmath}
\usepackage{amsthm}
\usepackage{gensymb}
\usepackage{graphicx,color}
\usepackage[dvipsnames]{xcolor}
\usepackage{graphicx}
\usepackage{mathtools}
\usepackage{amssymb,color}
\allowdisplaybreaks
\usepackage{subcaption}
\usepackage{cite}
\usepackage{blindtext}
\usepackage{tcolorbox}
\usepackage{colortbl}

\usepackage{xcolor}
\usepackage{lipsum}
\usepackage{algorithm}
\usepackage{algorithmic}
\usepackage{array} 
\usepackage{makecell}
\usepackage{tabularx}
\usepackage{bm}
\usepackage{siunitx, soul}

\newtheorem{rem}{Remark}
\usepackage{booktabs}

\usepackage[shortlabels]{enumitem}

\usepackage[font=scriptsize,labelfont=bf]{caption}

\newcommand{\tr}{\text{Tr}}
 
\newcommand{\thr}{\text{th}}
 
\makeatother
\bstctlcite{IEEEexample:BSTcontrol}
\newcommand{\qu}{\mathbf{u}}
\newcommand{\qh}{\mathbf{h}}
\newcommand{\qa}{\mathbf{a}}
\newcommand{\qe}{\mathbf{e}}
\newcommand{\qb}{\mathbf{b}}

\newcommand{\qg}{\mathbf{g}}
\newcommand{\qx}{\mathbf{x}}
\newcommand{\qv}{\mathbf{v}}
\newcommand{\qD}{\mathbf{D}}
\newcommand{\qE}{\mathbf{E}}
\newcommand{\qf}{\mathbf{f}}

\newcommand{\qH}{\mathbf{H}}
\newcommand{\qI}{\mathbf{I}}

\newcommand{\qV}{\mathbf{V}}
\newcommand{\qA}{\mathbf{A}}
\newcommand{\qW}{\mathbf{W}}
\newcommand{\qw}{\mathbf{w}}
\newcommand{\qS}{\mathbf{S}}
\newcommand{\qB}{\mathbf{B}}

\usepackage{enumitem}
\usepackage{multirow}

\renewcommand{\arraystretch}{1.3} 

\makeatother
\bstctlcite{IEEEexample:BSTcontrol}
\definecolor{LightGray}{gray}{0.9}
\definecolor{MediumGray}{gray}{0.5}
\definecolor{DarkGray}{gray}{0.2}

\begin{document}
\bstctlcite{IEEEexample:BSTcontrol}

\title{Downlink Beamforming for Cell-Free ISAC: A Fast Complex Oblique Manifold Approach}

\author{Shayan Zargari, Diluka Galappaththige, \IEEEmembership{Member, IEEE},  Chintha Tellambura, \IEEEmembership{Fellow, IEEE}, and Geoffrey Ye Li, \IEEEmembership{Fellow, IEEE,}
\thanks{S. Zargari, D. Galappaththige, and C. Tellambura with the Department of Electrical and Computer Engineering, University of Alberta, Edmonton, AB, T6G 1H9, Canada (e-mail: \{zargari, diluka.lg, ct4\}@ualberta.ca). \\
\indent G. Y. Li is with the ITP Lab, the Department of Electrical and Electronic Engineering, Imperial College London, SW7 2BX London, U.K.(e-mail: geoffrey.li@imperial.ac.uk).}
\vspace{-8mm}
}

\maketitle

\begin{abstract} 
Cell-free integrated sensing and communication (CF-ISAC) systems are just emerging as an interesting technique for future communications. Such a system comprises several multiple-antenna access points (APs), serving multiple single-antenna communication users and sensing targets.  However, efficient beamforming designs that achieve high precision and robust performance in densely populated networks are lacking. This paper proposes a new beamforming algorithm by exploiting the inherent Riemannian manifold structure. The aim is to maximize the communication sum rate while satisfying sensing beampattern gains and per AP transmit power constraints. To address this constrained optimization problem, a highly efficient augmented Lagrangian model-based iterative manifold optimization for CF-ISAC (ALMCI) algorithm is developed. This algorithm exploits the geometry of the proposed problem and uses a complex oblique manifold. Conventional convex-concave procedure (CCPA)  and multidimensional complex quadratic transform (MCQT)-CSA algorithms are also developed as comparative benchmarks. The  ALMCI algorithm significantly outperforms both of these.  For example, with \num{16} APs having \num{12} antennas and \qty{30}{\dB m} transmit power each, our proposed ALMCI algorithm yields \qty{22.7}{\percent} and \qty{6.7}{\percent} sum rate gains over the CCPA and MCQT-CSA algorithms, respectively.  In addition to improvement in communication capacity, the ALMCI algorithm achieves superior beamforming gains and reduced complexity.
\end{abstract}


\IEEEpeerreviewmaketitle
\section{Introduction}
\IEEEPARstart{C}{ell}-free (CF) integrated sensing and communication (ISAC) systems are emerging and receiving attention  \cite{3GPPISAC2024, Mao2023, Demirhan2023, demirhan2024cellfree, liu2024cooperative} recently. Their Applications include the Internet of Things (IoT), autonomous vehicles, indoor localization, and tracking \cite{3GPPISAC2024}. To achieve optimal sensing and communication performance,  transmit beamforming at the APs is critical \cite{Liu2022ISAC}.

CF-ISAC mitigates limitations of single AP or base station (BS)-based ISAC systems, such as obstructed observation angles that hinder target detection. By coordinating distributed APs, CF-ISAC systems enhance communication and sensing performance. This approach leverages the diversity gains from uncorrelated sensing observations at distributed receivers to improve sensing performance \cite{Mao2023, Demirhan2023, demirhan2024cellfree, liu2024cooperative}.

The performance of CF-ISAC critically depends on AP beamforming, which faces several technical challenges. APs must cooperatively serve users to meet rate demands while minimizing multi-user interference and enhancing beam gains for target detection. Traditional beamforming methods, like conjugate beamforming, may fail to address these issues, leading to poor sensing performance. Effective CF-ISAC beamforming techniques are needed to overcome these challenges \cite{Mao2023, Demirhan2023, demirhan2024cellfree, liu2024cooperative}.

\subsection{Previous Contributions}
CF-ISAC beamforming designs \cite{Mao2023, Demirhan2023, demirhan2024cellfree, liu2024cooperative} are at an embryonic stage. Existing studies exploit several classical techniques. Specifically, the successive convex approximation (SCA) algorithm in \cite{Mao2023} is based on minimizing the sensing beampattern gain matching mean-squared error. The AP beamforming in \cite{Demirhan2023} is designed based on a max-min fairness formulation to maximize the sensing signal-to-noise ratio (SNR) of a single target while adhering to a communication signal-to-interference-plus-noise ratio (SINR) constraint of multiple users. The semi-definite relaxation (SDR) approaches in \cite{demirhan2024cellfree} optimize beamforming while satisfying communication SINR restrictions. The beamforming and receive filtering in \cite{liu2024cooperative}  maximize the sum of sensing SINR. 
 
These prior algorithms can be classified into two types, namely (i) convex-concave procedure algorithms (CCPA) based on SDR and SCA  
(Section~\ref{sec_CCPA}), and (ii) fractional programming (FP)-based multidimensional complex quadratic transform (MCQT) algorithms
(Section~\ref{sec_FP_SCA}). 

    


However, approximations, linearization, and relaxation used in these works may incur performance losses. Furthermore, these algorithms may not scale well with densely populated networks and high computational demands. As user numbers or AP antenna arrays increase, they require more resources and longer execution times, needing more iterations to converge, which is problematic in dynamic, real-time environments.

\subsection{Manifold Optimization}

Manifold optimization (MO) differs from standard optimization techniques by exploiting the problem's inherent geometric structure. Instead of searching through the entire space, MO restricts the search to a manifold.
Thus, MO navigates the manifold's tangent spaces that locally resemble $\mathbb{R}^n$. These manifolds, such as lines, circles, planes, spheres, and matrix groups, have unique geometric properties. More details on MO can be found in \cite{liu2020simple, hu2020brief, zargari2024riemannian, boumal2023introduction, ring2012optimization} and references therein.

However, the existing CF-ISAC algorithms do not leverage the problem's geometric structure, which could enhance performance and convergence in high-dimensional scenarios if used properly. MO method is less sensitive to initialization and more scalable and therefore offers promising alternatives for large-scale, dynamic wireless networks \cite{liu2020simple,zargari2024riemannian}. They differ from conventional methods in gradient calculation, constraint handling, optimization path, and complexity \cite{liu2020simple, hu2020brief, zargari2024riemannian, boumal2023introduction}. While standard techniques may traverse infeasible regions needing correction, MO searches inherently stay on the manifold, ensuring a more direct and efficient trajectory \cite{ring2012optimization}.

MO-based algorithms have recently been developed for several communication and radar problems \cite{Chen2018, Zhou2017,  Chen2017, Zhong2022, zargari2024riemannian}. For instance, a data detection algorithm for massive multi-user MIMO systems has been developed in \cite{Chen2018}. In \cite{Zhou2017}, a Riemannian conjugate gradient (RCG)-based multi-cast beamforming algorithm is presented over a complex oblique manifold to maximize the minimum SINR across users in a coordinated multi-cell network.
The transmit beampattern design for MIMO radar systems in \cite{Zhong2022} converts the initial non-convex problem into an unconstrained quadratic Lagrange multiplier problem on a complex circle manifold. Reference \cite{Chen2017} solves the constant envelope precoding problem in multi-user massive MIMO systems on a complex circle manifold.
Our paper \cite{zargari2024riemannian} presents an innovative augmented Lagrangian manifold optimization (ALMO) transmit beamforming algorithm to maximize the communication sum rate while ensuring the ISAC sensing beam gain requirements. The previous MO methods are confined to specific system setups. The approach developed subsequently can readily accommodate any number of constraints, making it a more universal solution. 

\subsection{Our Contribution}
\begin{figure}[!t]
\centering 
\def\svgwidth{220pt} 
\fontsize{7}{7}\selectfont 
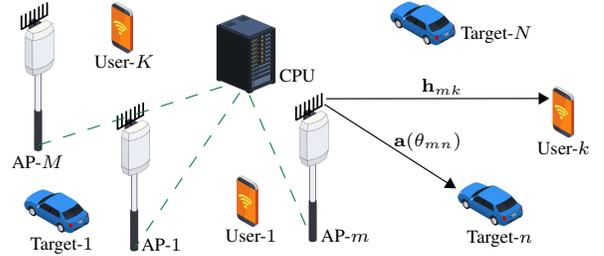\vspace{-0mm}  
\caption{A CF-ISAC system model.}  \label{fig_SystemModel}
\vspace{-2mm}
\end{figure}

The main contribution is developing an MO framework to improve CF-ISAC performance. Consider a generalized CF-ISAC system with multi-antenna APs, multiple communication users, and multiple sensing targets (Fig.~\ref{fig_SystemModel}). The goal is to design an optimal transmit beamforming strategy for the AP that maximizes the sum rate. This optimization is subject to constraints on the sensing beampattern gain and the per-AP transmit power limits. Since this problem is non-convex, a novel optimization algorithm is required. 

The main contributions are summarized as follows:
\begin{itemize}
\item  An innovative CF-ISAC  beamforming algorithm is proposed based on a search restricted to a complex oblique manifold and by augmenting the objective function with a  Lagrangian penalty factor \cite{liu2020simple}, which results in highly efficient augmented Lagrangian model (ALM)-based iterative MO for the CF-ISAC (ALMCI) algorithm.

\item Following our prior work on traditional ISAC \cite{zargari2024riemannian}, the proposed framework combines the dual objectives of maximizing the sum communication rate while meeting the sensing beampattern gain requirements. Nonetheless, this is the first implementation of this method to CF-ISAC, innovatively addressing the non-convex optimization challenges inherent in CF-ISAC beamforming.

\item The proposed ALMCI surpasses the state-of-the-art FP/SDR/SCA optimization techniques regarding communication sum rate and sensing beampattern gain. 
For instance, it respectively achieves gains of \qty{22.7}{\percent} and \qty{6.7}{\percent} in terms of sum rate for \num{16} APs with \num{12} antennas and \qty{30}{\dB m} transmit power each compared to CCPA and MCQT-SCA methods. Moreover, 
Furthermore, it also provides improved sensing beampattern gains while attenuating sidelobes.

\item Remarkably, such performance gains are achieved while reducing the computational complexity and algorithm execution time. For example, ALMCI is \num{13} and \num{27} times faster than CCPA and MCQT-SCA, respectively, for \num{4} APs with \num{16} antennas and at \qty{30}{\dB m} transmit power each. Moreover, the speed-up factor increases with the number of AP antennas. For instance, with 28  AP antennas, ALMCI is faster by   \num{60} and \num{34} times than the other two. 
Thus, the speed-up factor increases exponentially with the size of the network.

\item Reduced computational time and memory usage can lead to lower energy consumption. For example, ALMCI  uses \num{789} Megabytes (MB) fewer than  CCPA, a savings ratio of  \num{2724}.  Similarly, compared to MCQT-SCA, ALMCI saves \qty{11}{MB}, yielding a gain ratio of approximately \num{37}. Efficient computations, particularly in large-scale networks, result in less energy consumption. 
Thus, ALMCI may help to reduce the excessive resource consumption for large-scale CF-ISAC networks.
\end{itemize}

\textit{Notation}: $\mathbf{A}^{\rm{H}}$ and $\mathbf{A}^{\rm{T}}$ are the Hermitian (conjugate transpose) and the transpose of matrix $\mathbf{A}$. $\mathbf{I}_M$ denotes the $M$-by-$M$ identity matrix. The Euclidean norm,   absolute value, real part,  and expectation operators are denoted by $\|\cdot\|$ and $|\cdot|$, $\Re(\cdot)$,  and $\mathbb{E}\{\cdot\}$, respectively. Also, $\otimes$ denotes the Kronecker product. A circularly symmetric complex Gaussian (CSCG) random vector with mean $\boldsymbol{\mu}$ and covariance matrix $\mathbf{C}$ is denoted by $\sim \mathcal{C}\mathcal{N}(\boldsymbol{\mu},\,\mathbf{C})$. Besides, $\mathbb{C}^{M\times N}$ and ${\mathbb{R}^{M \times 1}}$ represent $M\times N$ dimensional complex matrices and $M\times 1$ dimensional real vectors, respectively. The $\text{unit}(\mathbf{a})$ function normalizes each element of vector $\mathbf{a}$. The Hadamard product $\mathbf{A} \circ \mathbf{B}$ denotes  element-wise multiplication, and $\text{clip}_{[a, b]} (x)$ limits $x$ to the interval $[a, b]$. And $\text{diag}(\cdot)$ is the diagonal matrix operator. Sets $\mathcal{K} \triangleq \{1,\ldots,K\}$, $\mathcal{N} \triangleq \{1,\ldots,N\}$, $\mathcal{M}_a \triangleq \{1,\ldots,M\}$, and $\mathcal{K}_k \triangleq \mathcal{K}\setminus{k}$ define indices, while $\mathcal{O}$ denotes big-O notation.

\section{System, Channel, and Signal Models}\label{Sec_system_modelA}
This section outlines the system, channel, and signal models. It describes the communication system architecture, the channel assumptions and mathematical representations, and the structure of the transmitted and received signals along with key parameters and variables.

\subsection{System and Channel Models}
As shown in Fig.~\ref{fig_SystemModel}, the CF-ISAC system comprises $M$ APs, each equipped with $L$ antennas, $K$ single-antenna communication users, and $N$ sensing targets. The AP has a uniform linear array (ULA) of antennas spaced at half-wavelength intervals, facilitating precise communication and sensing \cite{Zhenyao2023}.  The APs are connected to a central processing unit (CPU) via front-haul/back-haul links, allowing APs to jointly and coherently communicate with users while sensing targets. This can mitigate inter-user interference, providing a uniformly quality service to all communication users \cite{Ngo2017, Galappaththige2021, Galappaththige2024}.

Each AP has digital beamforming, with each antenna element connected to an independent RF chain, allowing simultaneous transmission of communication and sensing waveforms to serve both users and targets efficiently. A CPU coordinates joint design and processing at the APs, ensuring full synchronization across all APs \cite{Ngo2017}.

The communication environment is modeled using block flat-fading channels \cite{zargari2024riemannian}. Within each fading block, channel vector $\qh_{mk} \in \mathbb{C}^{L \times 1}$ represents the link from the $m$-th AP to the $k$-th user, and the vector $\qa(\theta_{mn}) \in \mathbb{C}^{L \times 1}$ represents the channel from the $m$-th AP to the $n$-th target. Specifically, communication channel is defined as
\begin{eqnarray}
\qh_{mk} = \zeta_{\mathbf{h}_{mk}}^{1/2} \tilde{\mathbf{h}}_{mk},~\forall k, m,
\end{eqnarray}
where $\zeta_{\mathbf{h}_{mk}}$ denotes the large-scale path-loss and remains constant over several coherence intervals, and  $\tilde{\mathbf{h}}_{mk} \sim \mathcal{CN}(\mathbf{0}, \mathbf{I}_{L})$ represents the small-scale Rayleigh fading and follows a complex normal distribution.

To model the sensing channels between the APs and the targets,  the echo signal representation in MIMO radar systems is adopted \cite{Zhenyao2023}. The line-of-sight (LoS) channels are thus represented using the transmit array steering vectors directed towards $\theta_{mn}$, where $\theta_{mn}$ is the $n$-th target's direction from the $m$-th AP for the $x$-axis of the coordinate system. Specifically, channel vector $\qa(\theta_{mn})$ is expressed as \cite{Zhenyao2023}:
\begin{equation}
\qa(\theta_{mn})\! = \! \frac{1}{\sqrt{L}} \!\left[1, e^{j\pi \sin(\theta_{mn})}, \ldots, e^{j\pi (L\!-\!1) \sin(\theta_{mn})} \right]^{\rm{T}}\!\!\!\!,~\forall n,m.
\end{equation}

\begin{rem}
The following assumptions are standard and widely used: 
\begin{itemize}
    \item[(i)] $\theta_{mn}$ is assumed to be pre-estimated known by the CPU for beamforming design, which can be obtained using past scanning \cite{Tsinos2021Joint, Zhenyao2023, Wu2018}, 

    \item[(ii)] A separate channel estimation phase is employed before communication and sensing to ensure CSI is available for beamforming design. 

    \item[(iii)] The APs and the users are connected via a controlled link, exchanging necessary commands \cite{positioningLTE}.
\end{itemize}

CF systems use the time division duplex (TDD) mode channel estimation. With this,  channel state information (CSI) can be readily estimated with algorithms, such as least squares (LS) and minimum mean-squared error (MMSE) estimators \cite{Marzettabook2016, Nayebi2018}.  Therefore,  the AP and users are assumed to have  CSI. This assumption is widely used \cite{demirhan2024cellfree, Demirhan2023, liu2024cooperative}.
\end{rem}

\subsection{Transmission Model}
The $m$-th AP transmit signal $\qx_m \in \mathbb{C}^{L\times 1}$  serves both communication and sensing and is  given by \cite{Zhao2022, zargari2024riemannian}
\begin{eqnarray}\label{eqn_tx_signal}
\qx_m = \sum_{k\in \mathcal{K}} \qv_{mk} q_{k},~\forall  m,
\end{eqnarray}
where $q_{k} \in \mathbb{C}$ is the data symbol for the $k$-th user with $\mathbb{E}\{\vert q_{k}\vert^2\} = 1$. The $m$-th AP beamforming vector for the $k$-th user, $\qv_{mk} \in \mathbb{C}^{L\times 1}$, helps to communicate with the $k$-th user while achieving improved sensing beampattern gain on target directions to maximize the radar's signal strength \cite{Zhao2022, Wang2022NOMA, Huang2022}. The received signal at the $k$-th user can be expressed as
\begin{align}\label{eqn_rx_user_k}
y_k
&=  \sum_{m\in \mathcal{M}_a} \qh_{mk}^{\rm{H}} \qv_{mk} q_k +  \sum_{m\in \mathcal{M}_a} \sum_{i\in \mathcal{K}_k} \qh_{mk}^{\rm{H}} \qv_{mi} q_i + n_k,~\forall k,
\end{align}
where $n_k \sim \mathcal{CN}(0, \sigma^2)$ denotes the $k$-th user additive white Gaussian noise (AWGN). To achieve the signal model in \eqref{eqn_rx_user_k},   clutter rejection techniques can mitigate reflected interference from the targets and surrounding environment \cite{Mark2010BOOK}.

\begin{rem}
While the APs can use separate beams for sensing, they also cause severe interference for communication users. To mitigate this, a dual-purpose beamforming strategy is employed to simultaneously serve communication and sensing purposes. It reduces the interference on communication users and maximizes the system’s efficiency by using a single beam to handle both functionalities \cite{Zhao2022, Wang2022NOMA, Huang2022}.
\end{rem}

Before proceeding to the problem formulation,  several useful notations are introduced.  First,  all the beamforming vectors of the $m$-th AP for $m\in\mathcal{M}_a$ are organized into a single vector $\mathbf{v}_m = [\mathbf{v}_{m1}^{\rm{T}}, \ldots, \mathbf{v}_{mK}^{\rm{T}}]^{\rm{T}} \in \mathbb{C}^{LK \times 1}$. Second, the matrix $\mathbf{V} = [\mathbf{v}_1, \ldots, \mathbf{v}_M] \in \mathbb{C}^{LK \times M}$, encapsulating beamforming vectors of all the APs for all the users, is constructed. 
Third, to effectively manage user and AP selections, selection matrices are introduced: $\mathbf{E}_k = [\mathbf{0}, \ldots, \mathbf{I}_L, \ldots, \mathbf{0}] \in \mathbb{R}^{L \times LK}$ for users, where the $k$-th block containing $L\times L$ identity matrix, i.e., $\mathbf{I}_L$, and $\mathbf{D} = \mathbf{I}_M \in \mathbb{R}^{M \times M}$ for APs.

\section{Joint Communication and Sensing}
This section focuses on deriving user communication rates and transmit beampattern gains for the targets. 

\subsection{Communication Performance}
Here, the rates at which users decode their data using the AP signal are specified. From \eqref{eqn_rx_user_k} and using the introduced notations, the SINR for the $k$-th user is expressed as
\begin{equation}\label{eqn_gamma}
\gamma_k=\frac{\vert \sum_{m\in \mathcal{M}_a} \qh_{mk}^{\rm{H}} \qE_k \mathbf{V} \qD_m \vert^2}{\sum_{i\in \mathcal{K}_k} \vert \sum_{m\in \mathcal{M}_a} \qh_{mk}^{\rm{H}} \qE_i \mathbf{V} \qD_m  \vert^2 + \sigma^2},~\forall k,
\end{equation}
where $\mathbf{D}_m$ represents the $m$-th column of matrix $\mathbf{D}$. From \eqref{eqn_gamma}, the  communication rate of the $k$-th user can be approximately represented as 
\begin{equation}
\mathcal{R}_k = \log_2(1+ \gamma_k), ~ \forall k.   
\end{equation}
The rate quantifies the effectiveness of data decoding by users.

\subsection{Sensing Performance}
The transmit beampattern gain is a widely used metric for evaluating sensing signal design \cite{Stoica2007}. Beampattern gains, which reflect the power distribution of the transmit signal across different sensing angles $\theta$, can enhance detection, recognition, and sensing accuracy  \cite{Stoica2007}.

The joint transmit beampattern gains can be used to achieve superior sensing outcomes, such as improved detection, range/Doppler/angle estimation, and tracking \cite{He2022}. This objective implies the joint processing of sensing signals across CPU's $M$ sensing receivers.
The sensing beampattern gain for the $n$-th target is given by \cite{He2022}
\begin{align}
p(\theta_{n}) &= \mathbb{E}\left\{\sum_{m\in \mathcal{M}_a}| \qa^{\rm{H}}(\theta_{mn}) \qx |^2 \right\} \\
&\!\!\!\!=\!\!\sum_{m\in \mathcal{M}_a} \sum_{k\in \mathcal{K}} \!\qa^{\rm{H}}(\theta_{mn}) \qE_k \mathbf{V} \qD_m (\qE_k \mathbf{V} \qD_m)^{\rm{H}} \qa(\theta_{mn}), \!~\forall n.\nonumber
\end{align}
As a result, this measure is designed specifically to meet the target sensing requirements. A uniformly distributed beampattern is most efficient if the target directions are unknown. Conversely, in applications where the potential directions of targets are approximately known, the beampattern gain should be maximized towards these directions, enhancing the efficiency of target detection and tracking \cite{Stoica2007}.

Other standard measures for evaluating sensing performance are transmitted beampattern mean-squared error (MSE) \cite{Liu2020} and sensing rate \cite{Zhenyao2023}. Specifically, the beampattern MSE minimizes the discrepancy between the sensing and ideal beampatterns. However, it requires the knowledge of the ideal beampattern, which may not be available. Conversely, sensing rate accounts for both transmit and receiver beampatterns, indicating how much environmental information can be extracted from a specific target's reflected signal. Yet, this demands the transmitter to operate in full-duplex mode, significantly impacting performance owing to self-interference. Thus, we employ beampattern gain as the viable performance metric \cite{Stoica2007, He2022}. Nevertheless, our approach can easily handle beampattern MSE and sensing rate.

Our ISAC design prioritizes two main metrics: (1)  the communication SINR and (2) the sensing beampattern gain. The former ensures high symbol detection accuracy and minimizes communication errors. On the other hand, sensing efficiency mainly depends on the beampattern gain, significantly affecting the target detection probability. A well-designed sensing beampattern gain helps achieve effective target recognition and overall sensing success \cite{Stoica2007}.

\subsection{Problem Formulation}
As mentioned above, the objective is to maximize the communication sum rate for users while meeting sensing targets and per-AP transmit power constraints. The problem can thus be stated  as follows:
\begin{subequations}
\begin{align}\label{P1}
\text{(P1)}:~& \max_{\mathbf{V}} \quad  \sum\nolimits_{k\in \mathcal{K}} \log_2\left(1 + \gamma_k \right),   \\
& \hspace{1mm}  \text{s.t.}  \quad   p(\theta_{n}) \geq \Gamma_{n}^{\thr}, ~\forall n, \label{P1_beamgain}\\
& \hspace{6mm} \sum\nolimits_{k\in \mathcal{K}} \Vert \qE_k \mathbf{V} \qD_m \Vert^2 \leq p_{\rm{max}}, ~\forall m.\label{P1_tx_pow}
\end{align}
\end{subequations}
where \eqref{P1_beamgain} guarantees the sensing beampattern gain over the threshold of the $n$-th target, $\Gamma_{n}^{\thr}$, and  \eqref{P1_tx_pow} indicates the  maximum allowable transmit power of the $m$-th AP, as $p_{\rm{max}}$. This optimization problem aligns with the envisioned capabilities of 6G ISAC systems \cite{Liu2022ISAC}.

\section{Optimal Beamforming} 
This section develops the MO solution to $\text{(P1)}$, utilizing FP and MO to determine optimal $\qv_{mk} \in \mathbb{C}^{L\times 1}$ \cite{liu2020simple, boumal2023introduction}. The key challenge of $\text{(P1)}$ is its non-convex objective function.

Due to the objective's non-convex sum-log terms, there is no known convex reformulation for the problem  $\text{(P1)}$. However, the Lagrangian dual transform can convert the objective to a sum-of-ratios form \cite[\textit{Theorem 3}]{Shen2018FPpart2}. By introducing auxiliary variables $\mu_k$ to replace SINR term $\gamma_k$ in \eqref{P1} such that $\mu_k \leq \gamma_k$, $\text{(P1)}$ is reformulated as \cite{Shen2018FPpart2}
\begin{subequations}\label{eqn_P3}
\begin{align}
\!\!\!\!\text{(P2)}:~& \max_{\mathbf{V}, \boldsymbol{\mu}} ~f(\mathbf{V}, \boldsymbol{\mu}) = \frac{1}{\ln(2)} \sum\nolimits_{k\in \mathcal{K}} \ln(1 + \mu_k)  \nonumber \label{obj_p3}\\
&\hspace{10mm} + \frac{1}{\ln(2)} \sum\nolimits_{k\in \mathcal{K}} \left( - \mu_k + \frac{(1 + \mu_k)\gamma_k}{1 + \gamma_k} \right) ,  \\
& \hspace{1mm}  \text{s.t.}  \quad   \eqref{P1_beamgain}-\eqref{P1_tx_pow},
\end{align}
\end{subequations}
where $\boldsymbol{\mu} = [\mu_1, \dots, \mu_K]$ represents the vector of auxiliary variables introduced by FP. Revised $\text{(P2)}$ embodies a dual-phase optimization challenge, comprising (i) an outer optimization of $\mathbf{V}$ with a static $\boldsymbol{\mu}$ and (ii) an inner optimization of $\boldsymbol{\mu}$ with a static $\mathbf{V}$. To address $\text{(P2)}$,  an iterative strategy is deployed, alternately refining $\mathbf{V}$ and $\boldsymbol{\mu}$ until the objective function stabilizes.

Note that the original problem is recast into an equivalent form, in which $\mathbf{V}$ resolves $\text{(P1)}$ if and only if it also solves $\text{(P2)}$ \cite[\textit{Theorem 3}]{Shen2018FPpart2}. Moreover, it is established that the optimal objective values of $\text{(P1)}$ and $\text{(P2)}$ are identical \cite{Shen2018FPpart2}.

\subsection{Optimization of $\boldsymbol{\mu}$ with Fixed $\mathbf{V}$}
The iterative optimization process first updates auxiliary variable $\boldsymbol{\mu}$ based on the previously determined values of $\mathbf{V}$. Notably, function $f(\mathbf{V}, \boldsymbol{\mu})$ becomes concave and differentiable over $\boldsymbol{\mu}$ when $\mathbf{V}$ remains constant, allowing for an optimal determination of $\boldsymbol{\mu}$ by setting $\frac{\partial f(\mathbf{V}, \boldsymbol{\mu})}{\partial \mu_k}=0$ for each component. The resulting update rule is thus given as \cite{Shen2018}
\begin{equation}\label{FP_uprule}
\mu_k^* =\gamma_k,~\forall k.
\end{equation}
For given $\boldsymbol{\mu}$, one can  eliminate the constant terms with respect to $\mathbf{V}$ in $\text{(P2)}$ to further simplify the optimization as
\begin{subequations}
\begin{align}\label{P4}
\!\!\!\!\text{(P3)}:\!\!~& \max_{\mathbf{V}} \!\!\!\!\! \quad  \sum_{k\in \mathcal{K}} \frac{\tilde{\mu}_k \vert  \sum_{m\in \mathcal{M}_a} \qh_{mk}^{\rm{H}} \qE_k \mathbf{V} \qD_m \vert^2}{\sum_{i\in \mathcal{K}} \vert \sum_{m\in \mathcal{M}_a} \qh_{mk}^{\rm{H}} \qE_i \mathbf{V} \qD_m \vert^2 + \sigma^2},\!\!  \\
&  \hspace{1mm} \text{s.t.}  \quad   \eqref{P1_beamgain}-\eqref{P1_tx_pow}, 
\end{align}
\end{subequations}
where $\tilde{\mu}_k = 1+\mu_k$ for $k \in \mathcal{K}$. It is emphasized that $\text{(P3)}$ and original problem $\text{(P1)}$ are equivalence and this transformation ensures no  performance loss  \cite{zargari2024riemannian}.

\begin{rem}\label{rem_equivalance}
The equivalence between $\text{(P1)}$ and $\text{(P3)}$ is established as follows: Upon substituting $\boldsymbol{\mu}^*$ back into $f(\mathbf{V}, \boldsymbol{\mu})$, the sum-of-logarithms objective function in $\text{(P1)}$, i.e., $\sum_{k\in \mathcal{K}} \log_2\left(1 +  \gamma_k \right)$ can be accurately recovered, affirming that $\text{(P2)}$ is equivalent to  $\text{(P1)}$ \cite{Shen2018, Shen2018FPpart2}. In \eqref{eqn_P3}, the sole term that depends on $\mathbf{V}$ is $\sum_{k\in \mathcal{K}} \frac{(1 + \mu_k) \gamma_k}{1 + \gamma_k}$. We can thus eliminate the constant terms related to $\mathbf{V}$ while maintaining the original objective \cite{boyd2004convex}. Hence, the objectives and constraints in $\text{(P2)}$ and $\text{(P3)}$ are identical, indicating their equivalence. The above equivalences, i.e., $\text{(P1)}$ to $\text{(P2)}$ and $\text{(P2)}$ to $\text{(P3)}$, establish the equivalence between the initial problem $\text{(P1)}$ and the final version $\text{(P3)}$ \cite{Shen2018, Shen2018FPpart2, boyd2004convex}.
\end{rem}

\subsection{Constrained Optimization on a Manifold} 
The $\text{(P3)}$ solution yields the optimal AP beamforming vectors. However, $\text{(P3)}$  has  two constraints. Among these, (8c) can be utilized to construct a suitable manifold for solving $\text{(P3)}$. Nonetheless, searching solely on this manifold is insufficient because constraint (8b) must also be satisfied. Note that unconstrained searches on manifolds are highly efficient, as demonstrated by recent works \cite{Chen2018, Zhou2017, Chen2017}. Consequently, we develop two key steps:  identifying the appropriate manifold and devising a penalty method to accommodate constraint (8b). These transform $\text{(P3)}$  into an unconstrained search on a manifold, streamlining the optimization process.

\subsubsection{\textbf{Choosing the right Manifold}}Usefully,  the power constraint is normalized such that the total power meets $\sum_{k\in \mathcal{K}} \Vert \qE_k \mathbf{V} \qD_m \Vert^2 \leq 1$. Each beamforming vector is extended by introducing a modified vector $\tilde{\mathbf{v}}_m = [\tilde{\mathbf{v}}^{\rm{T}}_{m1}, \ldots, \tilde{\mathbf{v}}^{\rm{T}}_{mK}]^{\rm{T}} \in \mathbb{C}^{(L+1)K \times 1}$, with each $\tilde{\mathbf{v}}_{mk} = [\mathbf{v}^{\rm{T}}_{mk}, z_k]^{\rm{T}} \in \mathbb{C}^{(L+1) \times 1}$ incorporating an additional element $z_k$ from the auxiliary vector $\mathbf{z} = [z_1, \ldots, z_K]$. This yields  the expanded matrix $\tilde{\mathbf{V}}=[\tilde{\mathbf{v}}_1, \ldots, \tilde{\mathbf{v}}_M] \in \mathbb{C}^{(L+1)K \times M}$.

These modifications ensure that the power normalization condition $\sum_{k\in \mathcal{K}} \Vert \tilde{\qE}_k \tilde{\mathbf{V}} \qD_m \Vert^2 = 1$ is satisfied, where $\tilde{\qE}_k=[\mathbf{0},\ldots, \mathbf{I}_L,\ldots, \mathbf{0}] \in \mathbb{R}^{(L+1) \times (L+1)K}$. A complex oblique manifold is defined  as follows:
\begin{equation}\label{eqn_M}
\mathcal{M} = \left\{ \mathbf{\tilde{V}} \in \mathbb{C}^{(L+1) K \times M} \:|\:  \Vert\tilde{\mathbf{V}}_{:1} \Vert^2 = \cdots= \Vert\tilde{\mathbf{V}}_{:M}\Vert^2  = 1 \right\}.
\end{equation}
Consequently, $\text{(P3)}$ is transformed into a constrained optimization problem on $\mathcal{M}$, yielding  $\text{(P4)}$,
\begin{figure*}[!t]
\begin{subequations}
\begin{align}\label{P5}
\text{(P4)}:~& \min_{\mathbf{\tilde{V}} \in \mathcal{M} } ~ \hat{f}(\mathbf{\tilde{V}} ) = -\sum_{k\in \mathcal{K}} \frac{\tilde{\mu}_k \vert \sum_{m\in \mathcal{M}_a} \mathbf{\hat{h}}_{mk}^{\rm{H}} \tilde{\qE}_k \tilde{\mathbf{V}} \qD_m  \vert^2}{\sum_{i\in \mathcal{K}} \vert \sum_{m\in \mathcal{M}_a} \mathbf{\hat{h}}_{mk}^{\rm{H}} \tilde{\qE}_i \tilde{\mathbf{V}} \qD_m \vert^2 + \sigma^2} ,  \\ 
&  \hspace{1mm} \text{s.t.}  \quad   \hat{g}_n(\mathbf{\tilde{V}} ) \!=\!\Gamma_{n}^{\thr} -\sum\nolimits_{m\in \mathcal{M}_a} \sum\nolimits_{k\in \mathcal{K}} \hat{\qa}^{\rm{H}}(\theta_{mn}) \tilde{\qE}_k \tilde{\mathbf{V}}  \qD_m (\tilde{\qE}_k \tilde{\mathbf{V}} \qD_m)^{\rm{H}} \hat{\qa}(\theta_{mn}) \leq 0 ,~\forall n \label{eqn_P5_sens}
\end{align}
\end{subequations}

\vspace{-2mm}

\hrulefill

\vspace{-3mm}

\end{figure*}
where $\mathbf{\hat{h}}_{mk} = \sqrt{p_{\rm{max}}}[\mathbf{h}_{mk}, 0]$ and $\hat{\qa}(\theta_{mn}) = \sqrt{p_{\rm{max}}}[\qa(\theta_{mn}), 0]$ adjust the dimensionality and scaling, respectively. 

\subsubsection{\textbf{Dealing with constraint \eqref{eqn_P5_sens}}} In $\text{(P4)}$, $\mathcal{M}$ is a Riemannian manifold, where the functions $\hat{f}:\mathcal M \rightarrow \mathbb R $ and $\hat{g}_n :\mathcal M \rightarrow \mathbb R $ are both twice continuously differentiable. Constraint \eqref{eqn_P5_sens} introduces a significant challenge, as unconstrained search on $\mathcal{M}$ may not satisfy it.  The ALM is employed to address this challenge \cite{zargari2024riemannian, Birgin2014book}.

The main trick is to add a penalty term to the main objective function. The penalty term punishes the violations of constraints, and the penalty intensity increases gradually. Such adaptations ensure an efficient enforcement of constraints, facilitating the convergence to optimal solutions. Thus, for problem $\text{(P4)}$, the set of constraints is given by \eqref{eqn_P5_sens}; thus, whenever $\hat g_n(x) >0$ occurs, a penalty term should be added. There can be several ways to do that. Specifically, the augmented Lagrangian function is defined as \cite{zargari2024riemannian, Birgin2014book} 
\begin{equation}\label{Lag_penalty}
\!\!\!\!\mathcal{L}_\rho(\mathbf{\tilde{V}} , \boldsymbol{\lambda}) = \hat{f}(\mathbf{\tilde{V}} ) + \frac{\rho}{2} \sum_{n \in \mathcal{N}} \!\max\left\{0, \frac{\lambda_n}{\rho} + {\hat{g}_n(\mathbf{\tilde{V}} )}\right\}^2\!,
\end{equation}
where $\rho > 0$ serves as a penalty parameter and $\boldsymbol{\lambda} (\geq 0) \in \mathbb{R}^{N}$ is the vector of the Lagrange multipliers. For fixed $\boldsymbol{\lambda}$, $\rho$ regulates the severity of the penalty imposed for any violations of \eqref{eqn_P5_sens} \cite{Birgin2014book}. A high $\rho$ leads to numerical instabilities, whereas too low $\rho$ can fail to penalize constraint violations adequately. Penalty factor $\rho$ is initially set and altered in response to how \eqref{eqn_P5_sens} is satisfied. If the improvement exceeds a certain level, $\rho$ remains the same; otherwise, it increases to strictly impose \eqref{eqn_P5_sens}.   Conversely, $\boldsymbol{\lambda}$ determines the objective function's sensitivity to constraints. They function as a correction term, aiding convergence to the true optimal solution \cite{Birgin2014book}. They are initialized with small positive values and iteratively adjusted to gradually enforce constraints to achieve balance between the constraints and the objective.

The ALM employs a cyclical strategy where  $\mathbf{\tilde{V}}$ is optimized with fixed $\boldsymbol{\lambda}$, an unconstrained search on $\mathcal M$, followed by an update of $\boldsymbol{\lambda}$  \cite{Birgin2014book}. This approach is called ALMO \cite{liu2020simple}. Thus, problem $\text{(P4)}$  is recast as
\begin{subequations}
\begin{align}\label{P6}
\text{(P5)}:~& \min_{\tilde{\qV} \in \mathcal{M},\boldsymbol{\lambda}} \quad  \mathcal{L}_\rho(\tilde{\qV} , \boldsymbol{\lambda}).
\end{align}
\end{subequations}

Next, the detailed steps to solve $\text{(P5)}$ are outlined. 

\subsubsection{\textbf{The proposed ALMCI algorithm}}
This  minimizes smooth function $\mathcal{L}_\rho(\mathbf{\tilde{V}}, \boldsymbol{\lambda})$ over complex oblique manifold $\mathcal{M} \rightarrow \mathbb{R}$. 

The algorithm computes  the Riemannian gradient, $\mathrm{grad}_{\mathbf{\tilde{V}}_t} \mathcal{L}_\rho(\mathbf{\tilde{V}}, \boldsymbol{\lambda})$, at current point $\mathbf{\tilde{V}}_t$ \cite{liu2020simple, zargari2024riemannian}. Then, the next point, $\mathbf{\tilde{V}}_{t+1}$, is obtained via a retraction mapping $R_{\mathbf{\tilde{V}}_t}$, projecting search direction vector $\eta{\mathbf{\tilde{V}}_t}$ back onto $\mathcal{M}$. A vector transport function $\mathcal{T}_{\mathbf{\tilde{V}}_t \rightarrow \mathbf{\tilde{V}}_{t+1}}$ then transfers the conjugate search direction from $T_{\mathbf{\tilde{V}}_t}\mathcal{M}$ to $T_{\mathbf{\tilde{V}}_{t+1}}\mathcal{M}$, ensuring the manifold's curvature and geometric properties. The reader is referred to Fig.~2 in \cite{zargari2024riemannian} for a visual representation of the MO steps. The main steps are summarized as follows:

(a) \textbf{Riemannian gradient}:
This is obtained by orthogonally projecting Euclidean gradient $ \nabla_{\mathbf{\tilde{V}}_t} \mathcal{L}_\rho(\mathbf{\tilde{V}} , \boldsymbol{\lambda}) $ onto tangent space $T_{\mathbf{\tilde{V}}_t}\mathcal{M}$. At $\mathbf{\tilde{V}}_t$, it can be obtained  as
\begin{align}\label{proj_grdman}
{\rm{grad}}_{\mathbf{\tilde{V}}_t} \mathcal{L}_\rho(\mathbf{\tilde{V}}, \boldsymbol{\lambda}) &= \nabla_{\mathbf{\tilde{V}}_t} \mathcal{L}_\rho(\mathbf{\tilde{V}}, \boldsymbol{\lambda}) \nonumber\\&
- \Re\{\nabla_{\mathbf{\tilde{V}}_t} \mathcal{L}_\rho(\mathbf{\tilde{V}} , \boldsymbol{\lambda}) \circ \mathbf{\tilde{V}}^*_t\}\circ \mathbf{\tilde{V}}_t,  
\end{align}
where  
\begin{equation}
T_{\mathbf{\tilde{V}}_t}\mathcal{M} = \left\{ \mathbf{c} \in \mathbb{C}^{(L+1) K} \mid \Re\{\mathbf{c} \circ \mathbf{\mathbf{V}}^*_t\} = \mathbf{0}_{(L+1) K} \right\},
\end{equation}
and $\mathbf{c} \in \mathbb{C}^{(L+1) K}$. The Euclidean gradient term in \eqref{proj_grdman} is derived by differentiating \eqref{Lag_penalty}  and is given in \eqref{derivtive_eq}.
\begin{figure*}[!t]
\begin{align}  \label{derivtive_eq}
\nabla_{\mathbf{\tilde{V}}_t} \mathcal{L}_\rho(\mathbf{\tilde{V}} , \boldsymbol{\lambda}) & =  \sum\nolimits_{k\in \mathcal{K}} -\tilde{\mu}_k 
\Bigg( \frac{2 \sum_{m\in \mathcal{M}_a}\mathbf{\hat{h}}_{mk}^{\rm{H}}  \tilde{\qE}_{k} \mathbf{\tilde{V}}_t  \qD_m  \sum_{m\in \mathcal{M}_a}  \tilde{\qE}_{k}^{\rm{H}}  \mathbf{\hat{h}}_{mk}  \qD_m^{\rm{H}}   }{\sum_{j\in \mathcal{K}} \vert \sum_{m\in \mathcal{M}_a} \mathbf{\hat{h}}_{mk}^{\rm{H}} \tilde{\qE}_j \tilde{\mathbf{V}} \qD_m \vert^2 + \sigma_k^2}\nonumber
\\& -   \sum\nolimits_{i\in \mathcal{K}} \frac{2\vert \sum_{m\in \mathcal{M}_a} \mathbf{\hat{h}}_{mk}^{\rm{H}} \tilde{\qE}_k \tilde{\mathbf{V}} \qD_m  \vert^2 \sum_{m\in \mathcal{M}_a}\mathbf{\hat{h}}_{mk}^{\rm{H}}  \tilde{\qE}_{i} \mathbf{\tilde{V}}_t  \qD_m  \sum_{m\in \mathcal{M}_a}  \tilde{\qE}_{i}^{\rm{H}}  \mathbf{\hat{h}}_{mk}  \qD_m^{\rm{H}} }{\left(\sum_{j\in \mathcal{K}} \vert \sum_{m\in \mathcal{M}_a} \mathbf{\hat{h}}_{mk}^{\rm{H}} \tilde{\qE}_j \tilde{\mathbf{V}} \qD_m \vert^2 + \sigma_k^2 \right)^2} \Bigg) \nonumber \\
& - 2 \rho \sum\nolimits_{n \in \mathcal{N}}   \mathbf{1}_{\left\{\lambda_n + \frac{\hat{g}_n(\mathbf{\tilde{V}} )}{\rho}\right\}} \left(\frac{\lambda_n}{\rho} + {\hat{g}_n(\mathbf{\tilde{V}} )}\right)  
\sum\nolimits_{m\in \mathcal{M}_a} \sum\nolimits_{k\in \mathcal{K}}\tilde{\qE}_k^{\rm{H}} \hat{\qa}(\theta_{mn}) \left(\hat{\qa}^{\rm{H}}(\theta_{mn})\tilde{\qE}_k  \mathbf{\tilde{V}}_t\qD_m \qD_m^{\rm{H}} \right)
\end{align}	

\vspace{-1mm}

\hrulefill

\vspace{-3mm}

\end{figure*}

(b) \textbf{Search direction and mapping}:
By extending traditional EO techniques  using $ {\rm{grad}}_{\mathbf{\tilde{V}}_t} \mathcal{L}_\rho(\mathbf{\tilde{V}}, \boldsymbol{\lambda}) $, the update rule for the search direction is given by
\begin{equation}\label{search_direction}
\boldsymbol{\eta}_{t+1} = -{\rm{grad}}_{\mathbf{\tilde{V}}_{t+1}} \mathcal{L}_\rho(\mathbf{\tilde{V}} , \boldsymbol{\lambda}) + \beta_t \mathcal{T}_{\mathbf{\tilde{V}}_t \rightarrow \mathbf{\tilde{V}}_{t+1}}(\boldsymbol{\eta}_{t}),
\end{equation} 
where $ \boldsymbol{\eta}_{t} $ is the current search direction and $ \beta_t $ is computed using the Hestenes-Stiefel approach \cite{Shewchuk1994}. Here, $\mathcal{T}_{\mathbf{\tilde{V}}_t \rightarrow \mathbf{\tilde{V}}_{t+1}}(\boldsymbol{\eta}_{t})$ is mapping a vector from the tangent space at $ \mathbf{\tilde{V}}_t$, $T_{\mathbf{\tilde{V}}_t}\mathcal{M}$, to the tangent space at $ \mathbf{\tilde{V}}_{t+1}$, $ T_{\mathbf{\tilde{V}}_{t+1}}\mathcal{M}$, which is given by
\begin{equation}\label{transport_op}
\mathcal{T}_{\mathbf{\tilde{V}}_t \rightarrow \mathbf{\tilde{V}}_{t+1}}(\boldsymbol{\eta}_{t}) = \boldsymbol{\eta}_{t} - \Re\{\boldsymbol{\eta}_{t} \circ \mathbf{\tilde{V}}^*_{t+1}\} \circ \mathbf{\tilde{V}}_{t+1}.
\end{equation}
Note that since $ \boldsymbol{\eta}_{t} $ and $ \boldsymbol{\eta}_{t+1} $ belong to different tangent spaces, thereby a vector transport is needed to maintain the integrity of the manifold's geometry.

(c) \textbf{Retraction}:
Once search direction $ \boldsymbol{\eta}_{t} $ is established at $ \mathbf{\tilde{V}}_t $, a retraction is employed to smoothly map this direction back onto the manifold, determining next point $ \mathbf{\tilde{V}}_{t+1} $. This retraction process is defined as
\begin{equation}\label{retraction_mapping}
\mathcal{R}_{\mathbf{\tilde{V}}_t}(\alpha_t \boldsymbol{\eta}_{t}) = \text{unt}(\alpha_t \boldsymbol{\eta}_{t}),
\end{equation}
where $ \alpha_t $ represents the step size along direction $ \boldsymbol{\eta}_{t} $. Thus, MO proceeds iteratively to converge to an optimal point where the Riemannian gradient vanishes, as defined in \eqref{Lag_penalty}. 

\begin{algorithm}[!t]
\caption{ALMCI Algorithm} 
\begin{algorithmic}[1] \label{alg:ALMOIMO}
\STATE \textbf{Require}: $\mathcal{M}$,  $\hat{f}(\mathbf{\tilde{V}})$, $\{\hat{g}_n(\mathbf{\tilde{V}})\}_{n \in \mathcal{N}}$: $\mathcal{M} \rightarrow \mathbb{R}$.
\STATE \textbf{Initialization}: Initial point $\mathbf{\tilde{V}}_0 \in \mathcal{M}$, Lagrange multipliers $\boldsymbol{\lambda}^0 \in \mathbb{R}^{N}$, accuracy tolerance $\epsilon_{\min}$, convergence tolerance $\delta_1>0$, initial accuracy $\epsilon_0 > 0$, initial penalty factor $\rho_0$, reduction factors $\theta_\epsilon \in (0, 1)$ and $\theta_\rho > 1$, boundaries for the multipliers $\lambda^{\min}_{n}, \lambda^{\max}_{n} \in \mathbb{R}$ ensuring $\lambda^{\min}_{n} \leq \lambda^{\max}_{n}$, ratio $\tau \in (0, 1)$, a minimum acceptable distance $d_{\min}$, and final convergence tolerance $\delta_2>0$.
\STATE Set $t = 0$.
\WHILE{$\text{dist}(\hat{f}(\mathbf{\tilde{V}}_t), \hat{f}(\mathbf{\tilde{V}}_{t+1})) \geq \delta_2$}
\STATE Set $j = 0$.
    \WHILE{$\text{dist}(\mathbf{\tilde{V}}_j, \mathbf{\tilde{V}}_{j+1}) \geq d_{\min}$ or $\epsilon_j > \epsilon_{\min}$}
        \STATE Set $i = 0$.
        \STATE Calculate $\boldsymbol{\eta}_0 = -{\rm{grad}}_{\mathbf{\tilde{V}}_0} \mathcal{L}_\rho(\mathbf{\tilde{V}}, \boldsymbol{\lambda})$ as Riemannian gradient according to \eqref{proj_grdman}.
        \WHILE{$\|{\rm{grad}}_{\mathbf{\tilde{V}}_i} \mathcal{L}_\rho(\mathbf{\tilde{V}}, \boldsymbol{\lambda})\|_2 > \delta_1$}
            \STATE  Determine the Armijo backtracking line search step size $\alpha_i$ according to \cite{Shewchuk1994}.
            \STATE Update $\mathbf{\tilde{V}}_{i+1}$ using retraction $R_{\mathbf{\tilde{V}}_i}(\alpha_i\boldsymbol{\eta}_i)$ as described in \eqref{retraction_mapping}.
            \STATE Compute the Riemannian gradient at the new point ${\rm{grad}}_{\mathbf{\tilde{V}}_{i+1}} \mathcal{L}_\rho(\mathbf{\tilde{V}}, \boldsymbol{\lambda})$ according to \eqref{proj_grdman}.
            \STATE  Obtain the vector transport $\mathcal{T}_{\mathbf{\tilde{V}}_i \rightarrow \mathbf{\tilde{V}}_{i+1}}(\boldsymbol{\eta}_i)$ using \eqref{transport_op}.
            \STATE Compute the Hestenes-Stiefel parameter $\beta_i$ according to \cite{Shewchuk1994}.
            \STATE  Update the conjugate gradient direction based on \eqref{search_direction}.
            \STATE $i \leftarrow i + 1$;
        \ENDWHILE
        \STATE Update the Lagrange multiplier using \eqref{lagmult_1}.
        \STATE Set $\sigma_{n}^{j+1} = \max \left\{ {\hat{g}_n(\mathbf{\tilde{V}}_{j+1})}, -\frac{\lambda_{n}^{j+1}}{\rho_t} \right\}, \forall n$.
        \STATE Adjust the accuracy tolerance $\epsilon_{j+1} = \max \{\epsilon_{\min}, \theta_\epsilon \epsilon_j\}$.
        \IF{$j = 0$ or $\underset{n}{\max} \{|\sigma_{n}^{j+1}| \} \leq \tau \underset{n}{\max} \{ |\sigma_{n}^{j}| \}$}
            \STATE Maintain the current penalty value: $\rho_{j+1} = \rho_j$.
        \ELSE
            \STATE Increment the penalty value: $\rho_{j+1} = \theta_\rho \rho_j$.
        \ENDIF
        \STATE $j \leftarrow j + 1$
    \ENDWHILE
    \STATE Calculate $\gamma$ according to the update rule defined in \eqref{FP_uprule}.
    \STATE Obtain $\mathbf{\tilde{V}}^{(t+1)}$.
    \STATE $t \leftarrow t + 1$
\ENDWHILE
\STATE \textbf{Output}: $\mathbf{V}^* = \mathbf{\tilde{V}}^{*}(1:LK, M)$.
\end{algorithmic}
\end{algorithm}

(d) \textbf{Updating the Lagrange multipliers}:
Once  $\mathbf{\tilde{V}}$ is optimized on $\mathcal{M}$,  $\boldsymbol{\lambda}$  is updated to meet the constraints. These updates include clipping safeguards that bound the multipliers,  ensuring that updates effectively contribute to resolving constraint violations. The update rule for each Lagrange multiplier at iteration $t$ is specified by
\begin{equation}\label{lagmult_1}
\lambda_n^{t+1} = \text{clip}_{[\lambda^{\min}_{n}, \lambda^{\max}_{n}]}\left(\lambda_n^{t} + \rho_t \hat{g}_n(\mathbf{\tilde{V}}_{t+1})\right),~\forall n,
\end{equation}
where $\rho_t > 0$ is the positive penalty parameter at iteration $t$. The clipping process restricts each $\lambda_n^{t+1}$ to a predetermined range $[\lambda^{\min}_{n}, \lambda^{\max}_{n}]$, preventing unbounded growth and maintaining stability \cite{liu2020simple,zargari2024riemannian}. Moreover, the updates occur only when significant progress is made in reducing constraint violations, thus avoiding premature or overly aggressive adjustments that may cause instability. 

\begin{figure}[!t]\centering\vspace{-4mm}
\includegraphics[width=0.45\textwidth]{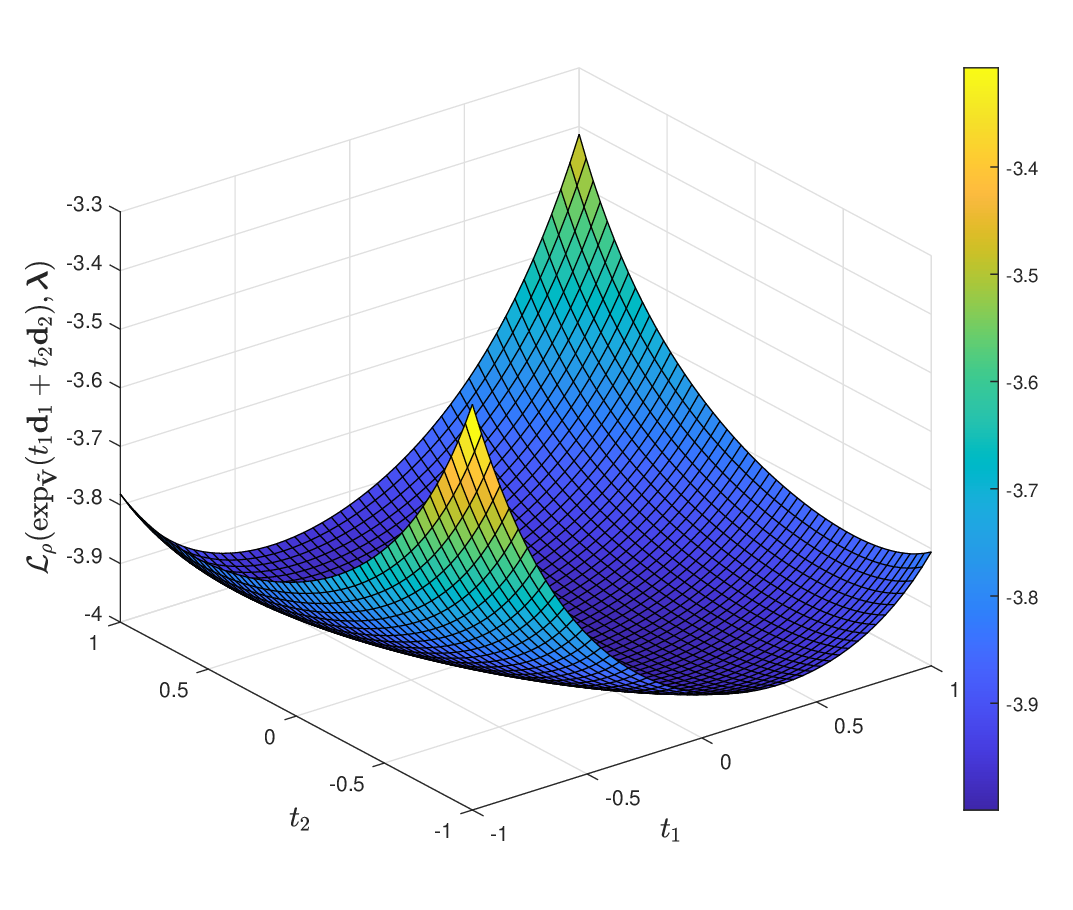}\vspace{-3mm}
\caption{The cost function of $\text{(P5)}$.   }
\label{fig_3D_surface}\vspace{-3mm}
\end{figure}

The overall process is given in Algorithm \ref{alg:ALMOIMO}. It converges to a global minimum of problem $(\text{P5})$ by generating an infinite sequence $\{\epsilon_j\}$ that converges to zero. At each iteration $j$, it produces a candidate solution $\mathbf{\tilde{V}}_{j+1}$ satisfying the condition $\mathcal{L}_{\rho_j}(\mathbf{\tilde{V}}_{j+1}, \boldsymbol{\lambda}_j) \leq \mathcal{L}_{\rho_j}(\mathbf{\tilde{V}}_j, \boldsymbol{\lambda}_j) + \epsilon_j$ \cite[Proposition 1]{zargari2024riemannian}. If the generated sequence has a limit point $\mathbf{\tilde{V}}^*$, this limit point is a global minimum. The algorithm ensures that the sequence of augmented Lagrangian values is non-increasing, thus bringing each iteration closer to the global solution. For first-order convergence, if the sequence has a limit point within the feasible set, then this limit point satisfies the Karush-Kuhn-Tucker (KKT) conditions of $(\text{P4})$ \cite[Proposition 2]{zargari2024riemannian}.



Fig.~\ref{fig_3D_surface} depicts the evaluation of a cost function of problem $\text{(P5)}$, where parameters $t_1$ and $t_2$ are varied across a two-dimensional grid spanning the interval $[-1,1]$. 
In addition, $\mathbf{d}_1$ and $\mathbf{d}_2$ are tangent vectors at point $\mathbf{\tilde{V}}$ on manifold $\mathcal{M}$. These vectors represent directions in which we can move from $\mathbf{\tilde{V}}$ while remaining on the manifold. They are obtained by orthogonally projecting the Euclidean gradient onto the tangent space at $\mathbf{\tilde{V}}$. This figure captures the concavity of function behavior $\mathcal{L}_\rho(\text{exp}_\mathbf{\tilde{V}}(t_1\mathbf{d}_1+t_2\mathbf{d}_2), \boldsymbol{\lambda})$ and highlights how the cost changes in response to movements along the manifold in the directions specified by these tangent vectors. This also serves as a visual guide to the algorithm's progress toward convergence, validating that each iteration moves closer to the solution by reducing the cost function value.

Note that our problem formulation and solution are easily adaptable to additional constraints on $\text{(P1)}$, such as minimum user rates, making it more versatile. 

\begin{table}[t]
\renewcommand{\arraystretch}{1.0}
\centering
\caption{Simulation and algorithm parameters.}\vspace{-1mm}
\label{table-notations}
\begin{tabular}{c c c c}    
\hline
\rowcolor{LightGray}
\textbf{Parameter}& \textbf{Value} & \textbf{Parameter}& \textbf{Value}\\  \hline \hline
$\sigma^2$ & \qty{-80}{\dB m}  & $\Gamma_{n}^{\thr}$  & \qty{20}{\dB m}  \\ 
$L$ & \num{16} & $p_{\rm{max}}$  & \qty{30}{\dB m} \\ 
$K$ & \num{2} & $\delta_1,\delta_2$  & \num{e-6}  \\ 
$N$ & \num{4} & $\nu$  & \num{2}  \\
$d_{\min}$ & \num{e-10} & $\epsilon_0$  & \num{e-3} \\
$\epsilon_{\min}$ & \num{e-6} & $\tau$  & \num{0.5}  \\
$\theta_\epsilon$ & \num{0.5} & $\theta_\rho$ & \num{0.25} \\
$\rho_0$ & \num{1} & $\{\lambda^{\min}_{n},\lambda^{\max}_{n}\}$ & \{0,100\} \\ \hline
\end{tabular}\vspace{-3mm}
\end{table}

The convergence of the algorithm for $\text{(P5)}$ is analyzed next, focusing on its ability to reach a global minimum, assuming optimal functionality of Algorithm \ref{alg:ALMOIMO}. The monotonic nature of the algorithm and an upper bound on the objective function ensure convergence, as detailed in \cite[\textit{Proposition 1}]{zargari2024riemannian}. Each iteration of the algorithm complies with the condition: $\mathcal{L}{\rho_t}(\mathbf{\tilde{V}}_{t+1}, \boldsymbol{\lambda}_t ) \leq \mathcal{L}{\rho_t}(\mathbf{\tilde{V}}_t, \boldsymbol{\lambda}_t) + \epsilon_t,$ confirming that $\mathbf{\tilde{V}}_{t+1}$ is a feasible global minimum. This sequence converges to a point $\mathbf{\tilde{V}}^*$ that also represents a global minimum of $\text{(P5)}$. The computational complexity for this process is dominated by the iterations of Algorithm \ref{alg:ALMOIMO}, yielding an approximate complexity of $\mathcal{O}(T(LMK + LMK^3))$, where $T$ denotes the total iterations.

\subsection{CCPA}\label{sec_CCPA}
This method is based on standard  SDR and SCA techniques. It is equivalent to algorithms in \cite{Mao2023, Demirhan2023, demirhan2024cellfree} and is developed as a comparative benchmark. 

By defining $\qW_k = \qw_k \qw_k^{\rm{H}}$ for $k \in\mathcal{K}$, where $\qw_k\in \mathbb{C}^{LM \times 1}$ is the beamformers of all APs, and $\qW_k\in \mathbb{C}^{LM \times LM}$ is a semi-definite matrix, i.e., $\qW_k \succeq 0$, with $\text{Rank}(\qW_k) = 1$, $\text{(P1)}$ is transformed into a standard semi-definite problem (SDP) by relaxing the rank one constraint, as outlined in $\text{(P6)}$ \eqref{P7} \cite{demirhan2024cellfree}. 
\begin{figure*}[!t]
\begin{subequations}\label{P7}
\begin{align}
\text{(P6)}:~& \max_{\{\qW_k\}_{k\in \mathcal{K}}} ~\sum\nolimits_{k\in \mathcal{K}}   \log_2\left(\sum\nolimits_{i\in \mathcal{K}} \qf_k^{\rm{H}} \qW_i \qf_k + \sigma^2\right) - \log_2\left( \sum\nolimits_{i \in\mathcal{K}_k} \qf_k^{\rm{H}} \qW^{(t)}_i \qf_k + \sigma^2\right)\nonumber\\
&\hspace{70mm} + \frac{\log_2(e)}{ 2^{\log_2\left( \sum\nolimits_{i \in\mathcal{K}_k} \qf_k^{\rm{H}} \qW^{(t)}_i \qf_k + \sigma^2\right)}} \sum\nolimits_{i \in\mathcal{K}_k} \qf_k^{\rm{H}} \left(\qW_i - \qW^{(t)}_i\right) \qf_k,  \\
&  \hspace{1mm} \text{s.t.}  \quad     \tr\left(\sum\nolimits_{m \in \mathcal{M}_a}\sum\nolimits_{k\in \mathcal{K}} \qD'_m \qg_n \qg_n^{\rm{H}} \qD'_m  \qW_k \right) \geq \Gamma_{n}^{\thr}, ~\forall n, \label{eq:sensing_constraint} \\
& \hspace{8mm} \tr\left(\sum\nolimits_{k\in \mathcal{K}} \qW_k\right) \leq p_{\rm{max}}, \label{eq:power_constraint} \\
& \hspace{8mm} \qW_k \succeq 0, ~ \forall k \label{eq:semidef_constraint}
\end{align}
\end{subequations}

\vspace{-3mm}

\hrulefill

\vspace{-3mm}

\end{figure*}
Additionally, the channel matrices for each AP to users and sensing targets are redefined as $\qf_k = [\qh_{1k}^{\mathrm{T}}, \ldots, \qh_{Mk}^{\mathrm{T}}]^{\mathrm{T}} \in \mathbb{C}^{LM \times 1}$ and $\qg_n = [\qa^{\mathrm{T}}(\theta_{1n}), \ldots, \qa^{\mathrm{T}}(\theta_{Mn})]^{\mathrm{T}} \in \mathbb{C}^{LM \times 1}$, respectively, and the matrix $\qD'_m \in \mathbb{R}^{LM \times LM}$ is a block diagonal matrix, composed of $M$ blocks, each corresponding to a AP. This configuration is represented by matrix $\text{diag}(\mathbf{0}, \dots, \mathbf{0}, \qI_L, \mathbf{0}, \dots, \mathbf{0})$, where $\qI_L$ is placed at the $m$-th block on the diagonal. This placement specifically activates the transmit antennas of the $m$-th AP. Note that the interior-point method can solve sub-problems using the SDP \cite{boyd2004convex}. Finally, to impose the relaxed rank one constraint, Gaussian randomization is utilized \cite{Qingqing}. 

According to \cite[Th. 3.12]{polik2010interior}, the order of complexity for a SDP problem with $m$ SDP constraints, which includes a $n \times n$ positive semi-definite matrix is given by $\mathcal{O}\left( \sqrt{n} \log\left(\frac{1}{\epsilon}\right) (mn^3 + m^2n^2 + m^3) \right)$ with $\epsilon > 0$. For problem \eqref{P7}, with $n = LM$ and $m = N + K + 1$, the approximate computational complexity for solving \eqref{P7} is $\mathcal{O}\left( \sqrt{LM} \log\left(\frac{1}{\epsilon}\right) \left((N+K)(LM)^{3}\right) \right).$

\subsection{MCQT-SCA}\label{sec_FP_SCA}
This method introduces auxiliary variables to simplify and solve problem $(\text{P1})$. Specifically, $(\text{P1})$ is first converted to $(\text{P2})$ given in \eqref{eqn_P3} with the optimal value of $\boldsymbol{\mu}$ based on \eqref{FP_uprule}. However, $(\text{P2})$ remains challenging due to its high-dimensional sum-of-fraction non-convex objective function and the sensing constraint. As it is not applicable to common FP methods, such as Dinkelbach’s algorithm, we employ the MCQT \cite{Shen2018}. MCQT extends scalar-form fractional programming to matrix-form, handling the non-convexity of high-dimensional fractions \cite{Shen2018}. By introducing another set of auxiliary variables $\boldsymbol{\zeta}=[\zeta_1,\ldots,\zeta_K]^{\rm{T}} \in \mathbb{C}^{K \times 1}$, $(\text{P2})$ is reformulated into $\text{(P7)}$ given in \eqref{P8},
\begin{figure*}[!t]
\begin{subequations}\label{P8}
\begin{align}
\text{(P7)}:~& \min_{\qV, \{\zeta_k\}_{k\in \mathcal{K}}, \{\mu_k\}_{k\in \mathcal{K}}} ~ \qV^{\rm{H}} \qA \qV - \Re \{2\qB^{\rm{H}}\qV\},  \\
&  \hspace{1mm} \text{s.t.}  \quad  g\left(\qV^{(t)}, \qa(\theta_{mn}))\right) + 2 \mathbf{S}^{\rm{T}}_{mk} \qa(\theta_{mn}) \left(\qa(\theta_{mn})^{\rm{H}} \mathbf{S}_{mk}\qV^{(t)}\right)
\left(\qV - \qV^{(t)}\right) \geq \Gamma_{n}^{\thr}, ~\forall n, \label{eq:sensing_constraint_fp} \\
& \hspace{8mm}   \qV^{\rm{H}} \qD''_m  \qV  \leq p_{\rm{max}}, ~\forall m \label{eq:power_constraint_FP}
\end{align}
\end{subequations}

\vspace{-3mm}

\hrulefill

\vspace{-3mm}

\end{figure*}
where 
\begin{equation}
g\left(\mathbf{V},\qa(\theta_{mn})\right)=\!\!\!\sum_{m \in \mathcal{M}_a} \!\sum_{k\in \mathcal{K}} \qa^{\rm{H}}(\theta_{mn}) \qS_{mk} \mathbf{V} (\qS_{mk} \mathbf{V})^{\rm{H}} \qa(\theta_{mn}).
\end{equation}
In addition, $\qA = \qI_{K}\otimes \sum_{k\in \mathcal{K}} \qf_k \zeta_k \zeta^{\rm{H}}_k \qf^{\rm{H}}_k$, $\qB = \left[\qb^{\rm{T}}_1,\ldots, \qb^{\rm{T}}_K \right]^{\rm{T}}$
with $\qb^{\rm{T}}_k=\sqrt{\tilde{\mu}_k}\qf_k  \zeta_k$, and $\qD''_m=\qI_{K}\otimes \{\qe_m\qe_m^{\rm{H}} \otimes \qI_{L} \}$ with $\qe_m \in \mathbb{R}^{M \times 1}$. In \eqref{eq:sensing_constraint_fp}, selection matrix $\qS_{mk}$ extracts $\qv_{mk}$ from $\qV$, and defined as $\mathbf{S}_{mk} = [
\mathbf{0}, \cdots, \mathbf{0}, \mathbf{I}_L, \mathbf{0}, \cdots, \mathbf{0}] \in \mathbb{R}^{L \times KLM}$, where identity matrix $\mathbf{I}_L$ starts at column index $(m-1)KL + (k-1)L + 1$ and spans $L$ columns. This places identity matrix $\mathbf{I}_L$ at the block corresponding to the position of $\qv_{mk}$ within vector $\qV$. Moreover,  the constraint in \eqref{eq:sensing_constraint_fp} is linearized using the SCA method \cite{liu2024cooperative}. Therefore,  this baseline is referred to as MCQT-SCA. The algorithm iteratively updates $\boldsymbol{\zeta}$ and $\qV$. Optimal $\boldsymbol{\zeta}$ can be obtained by
\begin{equation}
 \zeta^*_k = \frac{\sqrt{\mu_k}\qf^{\rm{H}}_k \qv_k}{\sum_{j\in \mathcal{K}} \qf^{\rm{H}}_k \mathbf{v}_{j} \left(\qf^{\rm{H}}_k \mathbf{v}_{j} \right)^{\rm{H}}  + \sigma^2},~\forall k,   
\end{equation}
where $\qv_k=[\qv^{\rm{T}}_{mk},\ldots,\qv^{\rm{T}}_{Mk}]^{\rm{T}} \in \mathbb{C}^{LM \times 1}$. Problem $(\text{P7})$ is a standard quadratically constrained quadratic program (QCQP) and can be efficiently solved by the interior-point method \cite{boyd2004convex}. In CF networks, the channels are high-dimensional, making matrix $\mathbf{A}$ considerably large $(MLK)$. Thus, inverting $\mathbf{A}$  has a  complexity of about $\mathcal{O}(M^3L^3K^3)$. This significant computational burden may impede the practical implementation of the precoding design using this method.

\begin{table*}[ht]
\centering
\caption{Comparison of convergence.}\vspace{-1mm}
\label{tab:convergence_data}
\sisetup{table-format=2.4}
\begin{tabular}{
S[table-format=1.0]
S[table-format=2.4] 
S[table-format=2.0] 
S[table-format=2.4] 
S[table-format=2.0] 
S[table-format=2.4] 
S[table-format=2.0] 
S[table-format=2.4] 
S[table-format=2.0]
}
\hline 
\rowcolor{LightGray}
 & 
\multicolumn{2}{c}{$L = \num{8}$} & 
\multicolumn{2}{c}{$L = \num{16}$} &
\multicolumn{2}{c}{$L = \num{8}$} & 
\multicolumn{2}{c}{$L = \num{16}$} \\
\rowcolor{LightGray}
& \multicolumn{2}{c}{$p_{\rm{max}} = \qty{25}{\dB m}$} & 
\multicolumn{2}{c}{$p_{\rm{max}} = \qty{25}{\dB m}$} & 
\multicolumn{2}{c}{$p_{\rm{max}} = \qty{30}{\dB m}$} & 
\multicolumn{2}{c}{$p_{\rm{max}} = \qty{30}{\dB m}$} \\
\hline 
\rowcolor{LightGray}
& {Sum rate (bps/Hz)} & {Iterations} & {Sum rate (bps/Hz)} & {Iterations} & {Sum rate (bps/Hz)} & {Iterations} & {Sum rate (bps/Hz)} & {Iterations} \\ \hline \hline
\rowcolor{white}
{ALMCI} & 26.7211 & 4 & 28.9003 & 4 & 30.0492 & 4 & 32.1721 & 4 \\
\hline 
\rowcolor{white}
{CCPA} &19.0686& 16 &  23.6679 & 19 & 23.2935 & 19 & 26.5908 & 15 \\
\hline 
\rowcolor{white}
{MCQT-SCA} & 25.0315 & 31 & 28.1321 & 31 & 27.0484 & 31 & 29.9797 & 31 \\ 
\noalign{\hrule height 1.1pt}
\end{tabular}
\vspace{-3mm}
\end{table*}


\section{Simulation Results}
This section presents the simulation results, demonstrating the performance of the proposed optimal transmit beamforming strategy. Detailed comparisons with benchmark methods are also included to illustrate the proposed approach's effectiveness and efficiency in various scenarios.

\subsection{Simulation Setup and Parameters}
Herein,  the configuration and parameters of the simulation are detailed. Unless specified otherwise, Table \ref{table-notations} lists the simulation parameters  \cite{Ngo2017, Galappaththige2021, Galappaththige2024}. The path-loss is modeled as $\zeta_{\mathbf{h}{mk}} = C_0 \left(\frac{d{mk}}{D_0}\right)^{-\nu}$, where $C_0=\qty{-30}{\dB}$ represents the path-loss at the reference distance $D_0 = \qty{1}{\m}$, $d_{mk}$ represents the individual link distance, and $\nu$ denotes the path-loss exponent.

A system with two APs is simulated. Users and sensing targets are uniformly distributed within a $500\times 500 \,\qty{}{\m^2}$ area. Each simulation result is averaged over  \num{e3} Monte Carlo trials.

To study  beampattern gains, AP-1 and AP-2 are placed at coordinates $(10, 10)$ and $(80, 80)$, respectively. From AP-1, the direction angles for users and sensing targets are $\{-40, 40\}\degree$ and $\{-80, -20, 20, 80\}\degree$, respectively, ensuring coverage of a broad angular range. Similarly, from AP-2, direction angles for users and sensing targets are $\{-50, 50\}\degree$ and $\{-70, -10, 10, 70\}\degree$, respectively. The circles and triangles denote communication users and sensing targets, respectively.

\subsection{Benchmark Schemes}
The following four benchmarks are used to evaluate the proposed MO approach comparatively. The first two, MMSE and zero-forcing (ZF) beamformers,  are known for their computational efficiency and ability to reduce interference \cite{Marzettabook2016}. We derived the last two based on previous works \cite{Mao2023, Demirhan2023, demirhan2024cellfree, liu2024cooperative}, which do not use the MO approach.

\begin{enumerate}
\item \textbf{Sensing-ignorant ZF beamforming:} This  beamformer is given by $\mathbf{V}_{\mathrm{ZF},m} = \qH_m(\qH_m^{\mathrm{H}}\qH_m)^{-1}$, where  $\qH_m = [\qh_{m1}, \ldots, \qh_{mK}] \in \mathbb{C}^{L \times K}$ for $k  \in\mathcal{K}$ and $m \in\mathcal{M}_a$ is the channel matrix from the $m$-th AP to all user. This only focuses solely on the communication users and eliminates interference from non-intended users.  

\item \textbf{Sensing-ignorant MMSE beamforming:} This beamformer can be represented as $\mathbf{V}_{\mathrm{MMSE},m} = \qH_m\left( \qH_m^{\mathrm{H}}\qH_m + \sigma^2 \mathbf{I}_{K} \right)^{-1} $ for $m \in\mathcal{M}_a$. This is also designed for communication only.  This minimizes the error between the estimated and actual signals and effectively balances the signal and noise.

\item \textbf{CCPA:} This is equivalent to the joint communication and sensing beamforming algorithm developed in  \cite{demirhan2024cellfree}. It solves $\text{(P1)}$ using conventional  SDR and SCA techniques (see Section~\ref{sec_CCPA}).  Thus, this benchmark allows our algorithm to be compared against the state-of-the-art. 

\item \textbf{MCQT-SCA:} This evaluates the beamforming method from \cite{liu2024cooperative}. It handles $\text{(P1)}$ with the MCQT and SCA approaches (see Section~\ref{sec_FP_SCA}) \cite{Shen2018}.
\end{enumerate}

\subsection{Convergence and Execution Times}
Table~\ref{tab:convergence_data} compares the convergence performance of the ALMCI, CCPA, and MCQT-SCA algorithms for different simulation setups. At $p_{\rm{max}} = \qty{25}{\dB m}$ with $L = \num{8}$ antennas, ALMCI stabilizes at \num{26.7211} (bps/Hz) sum rate by the fourth iteration. In contrast,  the other two converge at  \num{16} and \num{31} iterations but with lower sum rates. Increasing the transmit power or the number of AP antennas yields higher sum rates for all schemes. However, regardless of simulation settings, the proposed  ALMCI converges in \num{4} iterations. The other two algorithms need at least 15 and 31 iterations to converge yet achieve lower sum rates. These findings highlight the ability of  ALMCI to rapidly optimize performance in various environments.


\begin{figure}[!t]\centering\vspace{-4mm}
\includegraphics[width=0.45\textwidth]{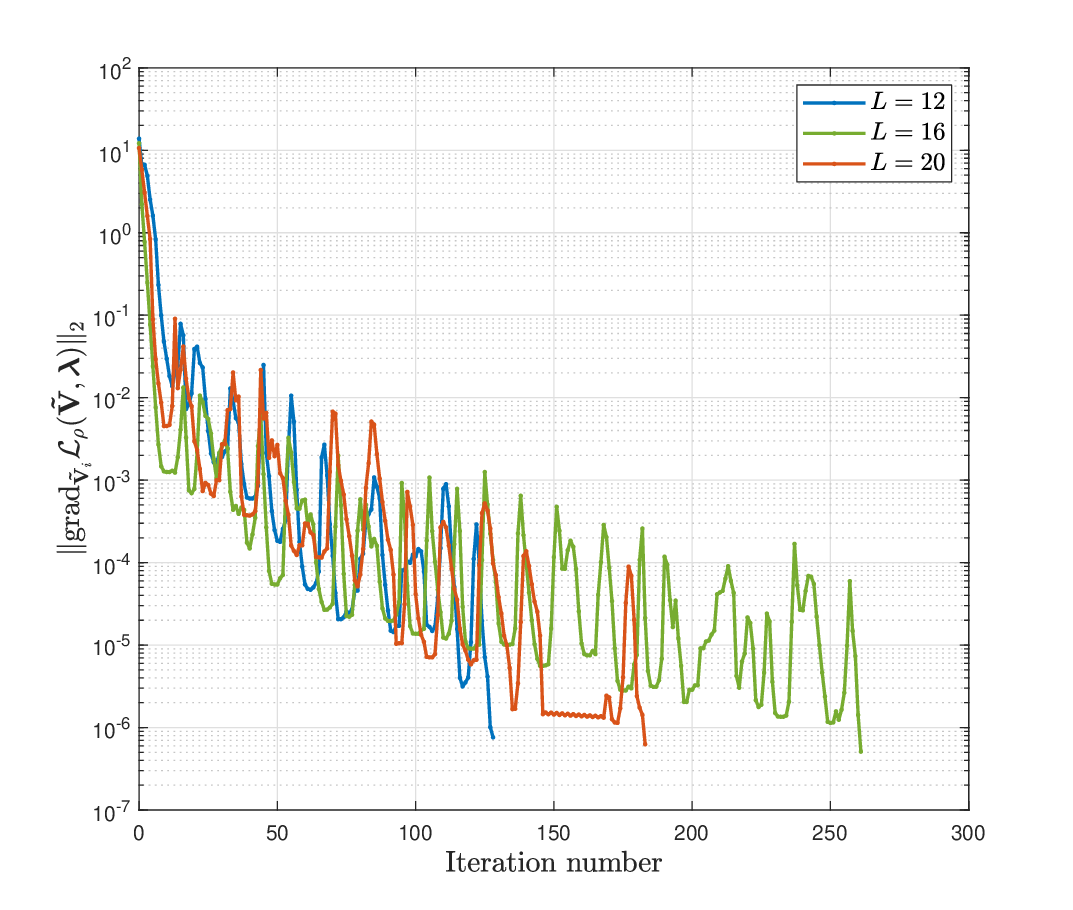}\vspace{-2mm} 
\caption{Convergence rate of the RCG with different numbers of AP antennas.}\vspace{-3mm}
\label{fig_convergence_CG}
\end{figure}

\begin{figure}[!t]\centering\vspace{-2mm}
\includegraphics[width=0.45\textwidth]{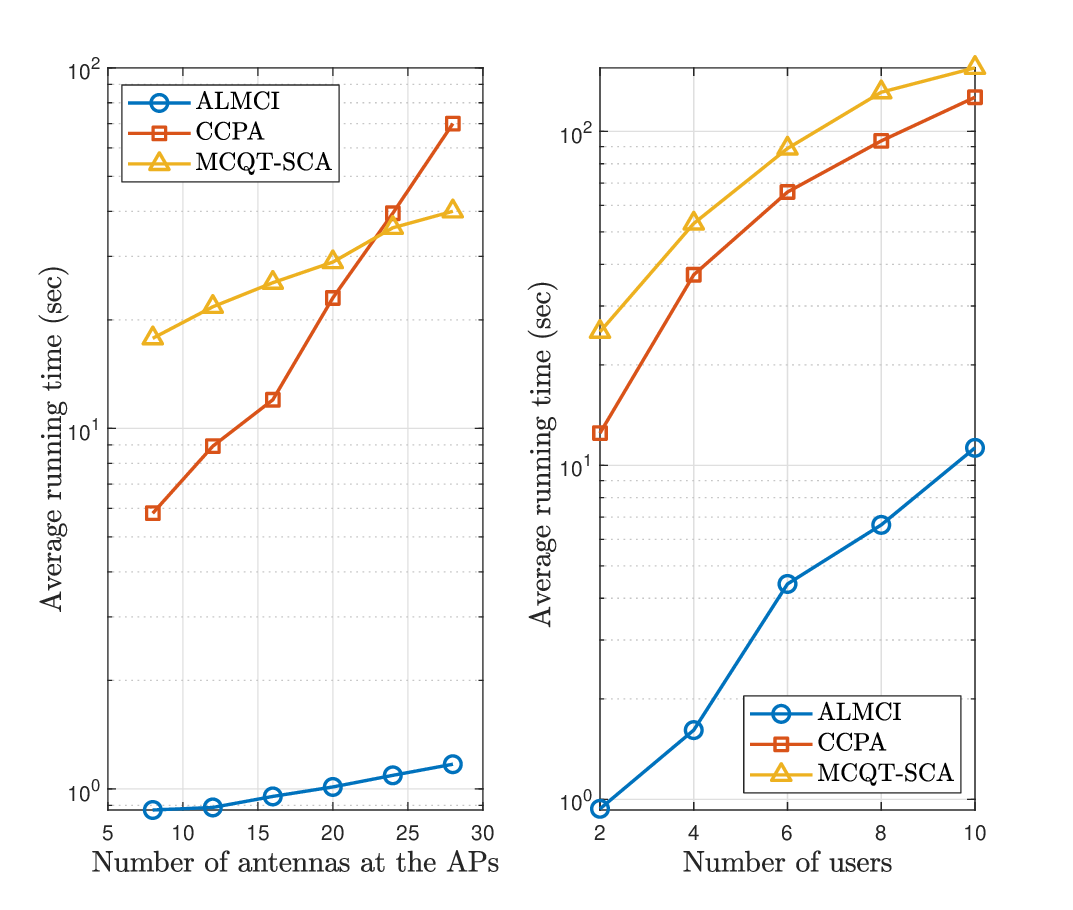} \vspace{-2mm}
\caption{Average running time versus number of users and AP antennas.}\vspace{-3mm}
\label{fig_time_combine}
\end{figure}

Fig.~\ref{fig_convergence_CG} illustrates the norm of the gradient of the Lagrangian function, $\|{\rm{grad}}_{\mathbf{\tilde{V}}i} \mathcal{L}\rho(\mathbf{\tilde{V}}, \boldsymbol{\lambda})\|_2$, as it evolves across iterations for various numbers of antennas. Initially, the gradient norm decreases rapidly for all configurations, indicating that the optimization algorithm quickly approaches the optimal regions characterized by lower gradient norms. As the iterations progress, this decrease becomes more gradual and fluctuations become more frequent. These fluctuations indicate the algorithm's step size and direction adjustment based on gradient guidance. Notably, the amplitude of these fluctuations becomes larger with higher values of $L$. This increase arises from the challenges of managing a higher-dimensional space, highlighting the impact of antenna numbers on the optimization's behavior.

\begin{table}[ht]
    \centering
    \caption{Comparison of memory usage.}
    \begin{tabular}{>{\raggedright}p{1.5cm} >{\centering}p{1.5cm} >{\centering}p{1.5cm} >{\centering\arraybackslash}p{1.5cm}}
        \hline
        \rowcolor{LightGray}
        \textbf{Method} & $M=\num{2}$ & $M=\num{3}$ & $M=\num{4}$ \\
        \hline \hline
         ALMCI    & \qty{0.29}{MB}   & \qty{20.59}{MB} & \qty{29.71}{MB} \\ \hline
        CCPA    & \qty{0.79}{GB}  & \qty{3.81}{GB}  & \qty{12.04}{GB} \\ \hline
         MCQT-SCA & \qty{10.74}{MB} & \qty{28.56}{MB} & \qty{37.20}{MB} \\
        \noalign{\hrule height 1.1pt}
    \end{tabular}
    \label{tab:memory_usage}
\end{table}

Fig.~\ref{fig_time_combine} plots the average running times for the ALMCI, CCPA, and MCQT-SCA algorithms, evaluating their performance across varying numbers of AP antennas and different numbers of users. The first graph reveals that the average running times for CCPA and MCQT-SCA increase with the antenna number. These trends reflect the algorithms' iterative nature and the computational intensity of the SCA/SDR process, increasing computational complexity as the antenna array expands. In contrast, ALMCI  exhibits a more modest and gradual increase, highlighting its computational efficiency and ability to exploit larger antenna arrays. For example, with \num{4} APs having \num{16} antennas and \qty{30}{\dB m} transmit power,  ALMCI  has \num{12} and \num{27} times faster algorithm execution time compared to the CCPA and MCQT-SCA algorithms, respectively.

The second graph shows the average run time as a function of the number of users.  All algorithms show a rising trend; however, CCPA and MCQT-SCA increase sharply, i.e., poor scalability with user growth. Because CCPA and MCQT-SCA are required to solve larger optimization problems and execute more complex matrix operations. Conversely, ALMCI demonstrates a slower increase in running time, indicating that its MO  approach scales more effectively and handles additional users more efficiently.

The ALMCI's reduced execution time and computational complexity stem from its search in a manifold with $(L+1)KM$ dimensions, compared to the larger Euclidean space of $MLKN$ dimensions used by conventional methods like CCPA and MCQT-SCA. This dimensionality reduction simplifies gradient computations, step size tuning, memory usage, and numerical stability, eliminating the need for explicit constraint handling and iterative feasibility checks. ALMCI's use of Riemannian gradients and manifold curvature provides precise updates and large steps, enabling fast convergence  \cite{liu2020simple,hu2020brief,boumal2023introduction,lee2006riemannian}. These features make it ideal for real-time ISAC systems in large and dynamic networks.

Table~\ref{tab:memory_usage} provides a comparison of memory usage across three different values of $M$ in megabytes (MB) and gigabytes (GB). As $M$ increases, the memory usage for all methods also increases. The CCPA method consistently uses significantly more memory than both ALMCI and MCQT-SCA while ALMCI uses the least memory. The difference in memory usage between CCPA and the other two methods is quite substantial, especially for higher values of $M$. Specifically, ALMCI achieves a memory savings of \qty{789.71}{MB} compared to CCPA, resulting in an impressive gain ratio of approximately \num{2724.86}. Similarly, when compared to MCQT-SCA, ALMCI saves \qty{10.45}{MB}, yielding a gain ratio of approximately \num{37.03}.

Methods requiring more memory (e.g., CCPA) will increase energy consumption since more power is required to maintain and access larger amounts of memory. The energy consumption of a chip is influenced by both dynamic and static power consumption \cite{chandrakasan1995minimizing}. Higher memory usage can increase dynamic power consumption due to more frequent memory access and static power consumption due to leakage currents in larger memory arrays. Increased energy consumption leads to high heat dissipation, which can impact the thermal management of the chip \cite{chandrakasan1995minimizing}. Effective cooling mechanisms might be needed, adding to the overall power budget and complexity of the chip design. Memory is a significant component of the overall cost of a chip. High memory requirements will increase the cost of materials. Methods like CCPA that require GB of memory could substantially increase the cost compared to methods like ALMCI or MCQT-SCA that require only MB of memory \cite{nguyen2023mcaimem}. Chips designed to support higher memory capacities may need to be larger in size, which can increase manufacturing costs. Additionally, more complex memory management circuitry may be needed, further adding to the design and production costs \cite{nguyen2023mcaimem}.



\begin{figure}[!t]\vspace{-4mm}
\centering
\begin{subfigure}[b]{0.45\textwidth}
\centering
\includegraphics[width=\textwidth]{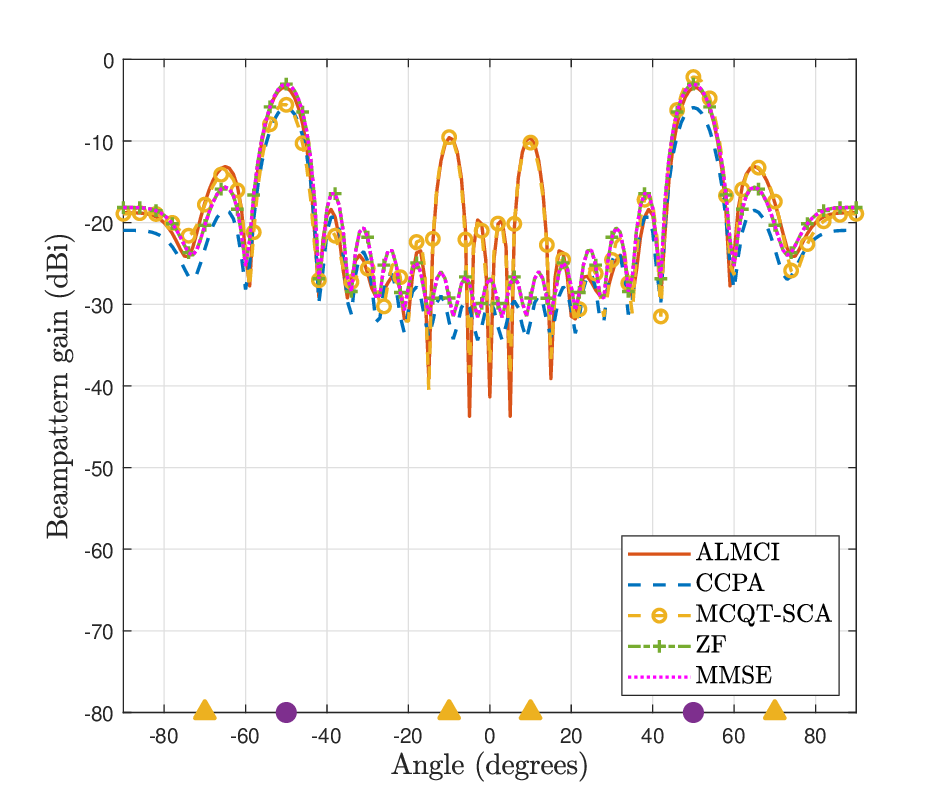}\vspace{-3mm}
\caption{AP-1}
\label{bamgain_antenna20_numUser2_AP1}
\end{subfigure}\vspace{-0mm}
\hfill  
\begin{subfigure}[b]{0.45\textwidth}
\centering
\includegraphics[width=\textwidth]{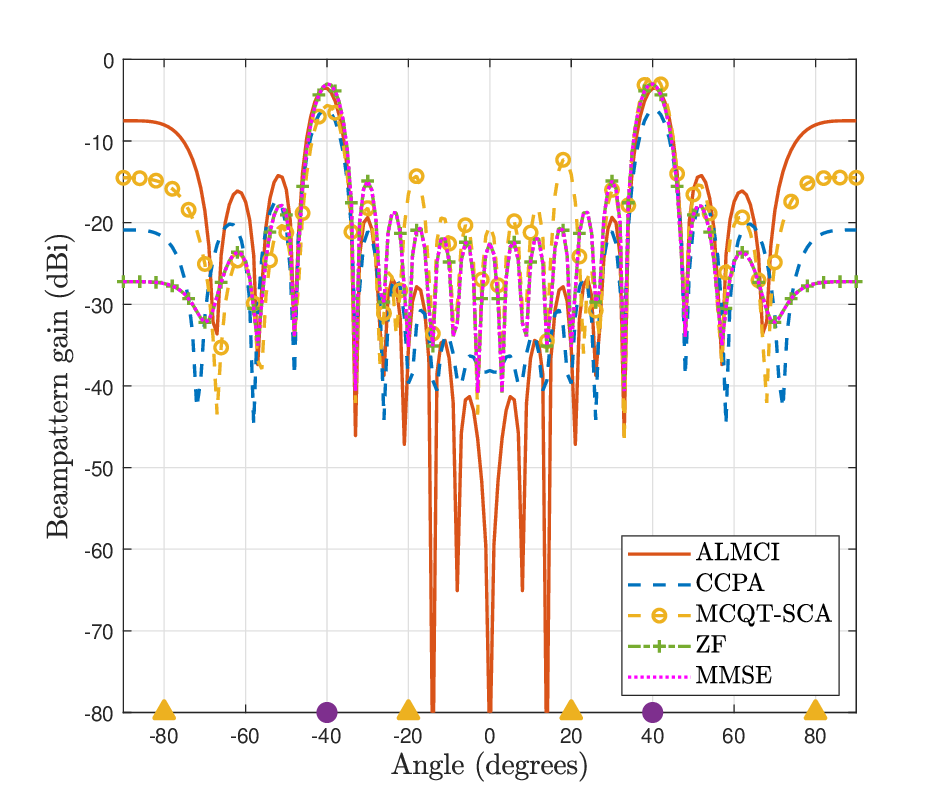}\vspace{-3mm}
\caption{AP-2}
\label{bamgain_antenna20_numUser2_AP2}
\end{subfigure}\vspace{-1mm}
\caption{Directional gain profiles for various algorithms over a \qty{\pm 90}{\degree} angular spread with $K=\num{2}$ user, $N=\num{4}$ targets, and $L=\num{20}$ antennas.}
\label{fig:combined1}\vspace{-3mm}
\end{figure}

\begin{figure}[!t]\vspace{-4mm}
\centering
\begin{subfigure}[b]{0.45\textwidth}
\centering
\includegraphics[width=\textwidth]{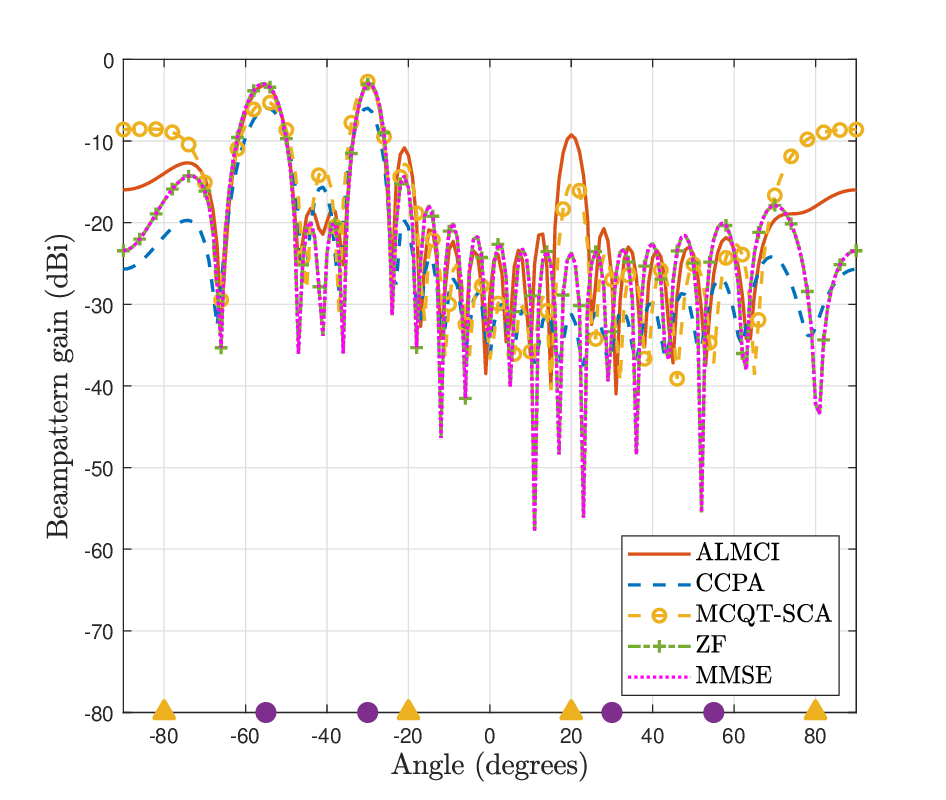}\vspace{-3mm}
\caption{AP-1}
\label{bamgain_antenna20_numUser4_AP1}
\end{subfigure}
\hfill   
\begin{subfigure}[b]{0.45\textwidth}
\centering
\includegraphics[width=\textwidth]{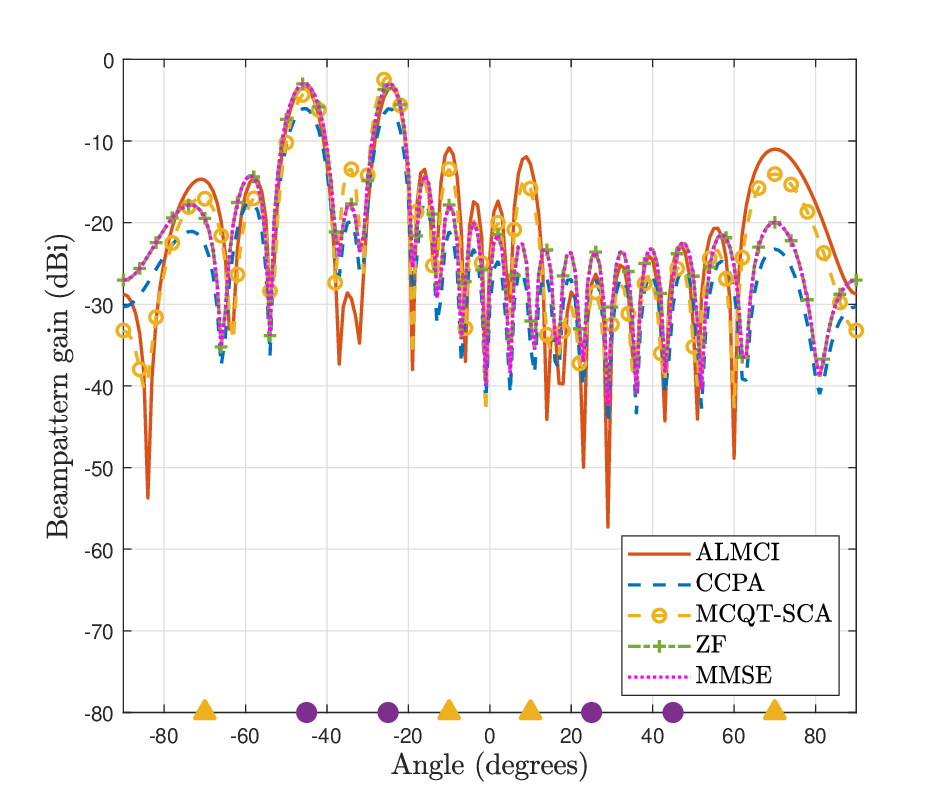}\vspace{-3mm}
\caption{AP-2}
\label{bamgain_antenna20_numUser4_AP2}
\end{subfigure}\vspace{-1mm}
\caption{Directional gain profiles for various algorithms over a \qty{\pm 90}{\degree} angular spread with $K=\num{4}$ user and $N=\num{4}$ targets, and $L=\num{20}$ antennas.}
\label{fig:combined2}\vspace{-3mm}
\end{figure}

\begin{figure}[!t]\vspace{-4mm}
\centering
\begin{subfigure}[b]{0.45\textwidth}
\centering
\includegraphics[width=\textwidth]{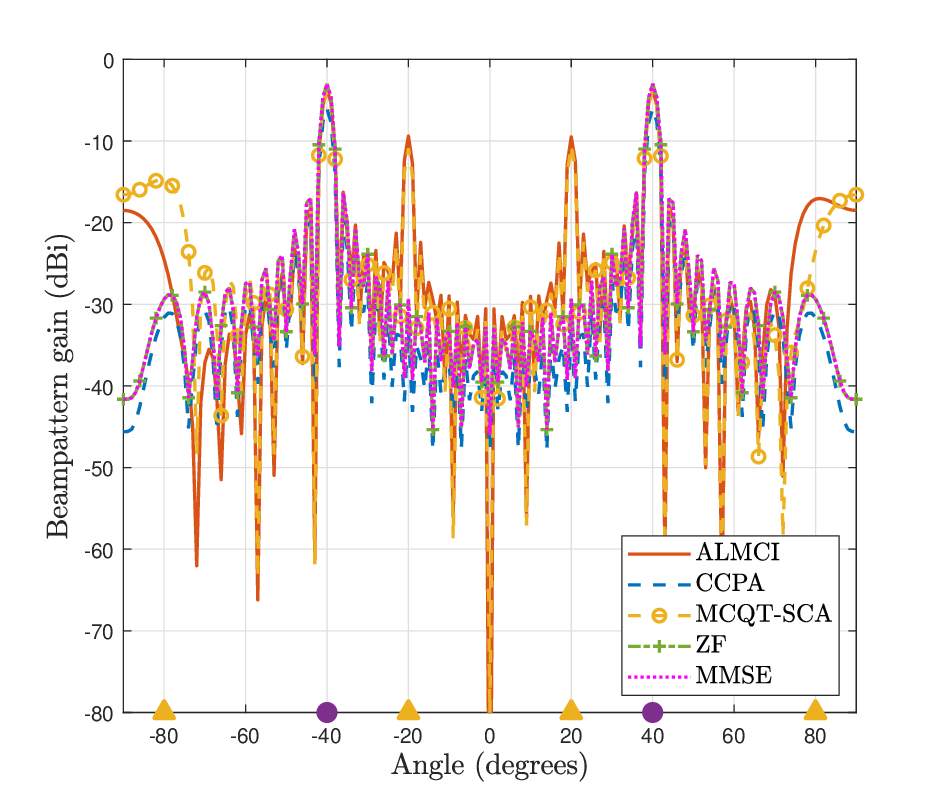}\vspace{-3mm}
\caption{AP-1}
\label{bamgain_antenna50_AP1} 
\end{subfigure}
\hfill  
\begin{subfigure}[b]{0.45\textwidth}
\centering
\includegraphics[width=\textwidth]{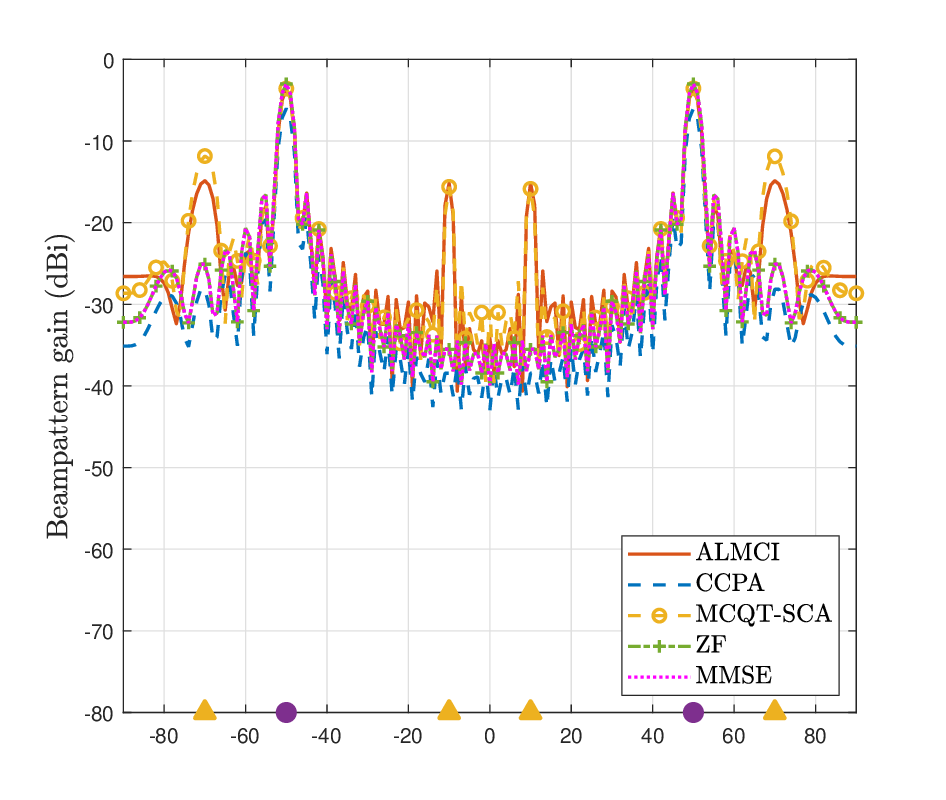}\vspace{-3mm}
\caption{AP-2}
\label{bamgain_antenna50_AP2}
\end{subfigure} \vspace{-1mm}
\caption{Directional gain profiles for various algorithms over a \qty{\pm 90}{\degree} angular spread with $K=\num{2}$ user and $N=\num{4}$ targets and $L=\num{50}$ antennas.}\vspace{-3mm}
\label{fig:combined5}
\end{figure}

\subsection{Sensing Performance}
Fig.~\ref{fig:combined1} plots the directional gain profiles with $K=\num{2}$ users and $N=\num{4}$ targets. The subfigures (a) and (b) are for  AP-1 and AP-2, respectively. The ALMCI algorithm designs the beam shape with high precision to match the user distribution and the sensing environment. It shows sharp, distinct peaks patterning a strong directional focus. This is expected as MO  exploits the geometric structure of the problem to achieve finer beam steering and width control, which minimizes interference and maximizes signal strength in the desired directions. The CCPA and MCQT-SCA algorithms, however,  demonstrate broader peaks. This beampattern spread may result from relaxing the non-convex problem into a convex one and the concomitant loss of optimality. The broader beams of  CCPA and MCQT-SCA may not be as effective in null-steering or avoiding interference as ALMCI, which impacts negatively when user channels are closely spaced. 

The conventional ZF/MMSE algorithms offer baselines for comparison. The ZF ideally creates deep nulls in non-intended directions. However, this inversion can lead to power amplification issues when dealing with channels that are near singular or when noise enhancement occurs. The MMSE approach introduces a compromise between inverting the channel and minimizing the power of the transmitted signal. Ultimately, the choice of algorithm depends on the specific requirements of the communication system or sensing application. The accuracy and low computational complexity of the ALMCI in beamforming are suitable for high-density scenarios with strict interference constraints. In contrast, ZF and MMSE may be preferred for their simplicity and reduced computational demands in less challenging environments.

Fig.~\ref{fig:combined2} showcases the directional gain with increased users $K=\num{4}$. The directional angles for users from AP-1 and AP-2  are selected as  $\{-55, -30, 30, 55\}\degree$ and $\{-45, -25, 25, 45\}\degree$, respectively. For AP-1, the ALMCI algorithm’s gain profile is characterized by its precision in peak alignment with user locations, indicating an effective concentration of signal power in the desired directions. This precise alignment can lead to enhanced signal reception for the intended users. Again, the CCPA and MCQT-SCA algorithms show broader peaks that result from their dependence on SCA/SDR optimizations. These broader beams can indicate a potential for spillover effects where signal power is distributed over a wider angular range, and is less efficient than ALMCI. The beamforming pattern of ZF/MMSE strikes a balance between interference reduction and gain maximization; their effectiveness is highly dependent on the channel conditions, and their performance degrades when users are closely spaced.

For AP-2, similar trends are observed although the magnitude and sharpness of the peaks appear slightly varied, reflecting the impact of the AP’s spatial positioning relative to users. This change suggests that environmental factors and user distribution considerably influence the algorithms’ performance. In both AP scenarios, more users introduce optimization challenges, such as a denser spatial user distribution and a more complex interference pattern. As ALMCI maintains high precision in its beamforming, it is also suitable for environments with high user density.

In Fig.~\ref{fig:combined5}, directional gain profiles of the algorithms are illustrated for the number of antennas  $L=\num{50}$. This increase in antennas affords finer control over beam shaping, clearly evidenced in the gain profiles for both AP-1 and AP-2. The ALMCI algorithm distinguishes itself, especially at AP-1, by delivering highly focused beam peaks. These sharp peaks indicate its precision in directing energy toward intended targets, significantly enhancing peak gains for sensing targets compared to those achieved by all other algorithms. Such precise beam direction could translate into improved spatial resolution for sensing applications and elevated data rates for communication users due to reduced inter-user interference.

Moreover, the ALMCI algorithm reduces sidelobes (i.e., interference) relative to other algorithms. Thus, its efficacy in minimizing potential interference is evident. Conversely, the gain profiles of the other two lack the same peak sharpness as ALMCI, suggesting a less targeted energy distribution. This indicates that CCPA and MCQT-SCA might not fully exploit the spatial multiplexing benefits of the antenna array. The ZF and MMSE algorithms align the beampattern gains toward the communication users without sensing the targets. Overall, expanding the antenna array magnifies the capabilities of all the algorithms. 

\subsection{Communication Performance}
\begin{figure}[!t]\centering\vspace{-3mm}
\includegraphics[width=0.45\textwidth]{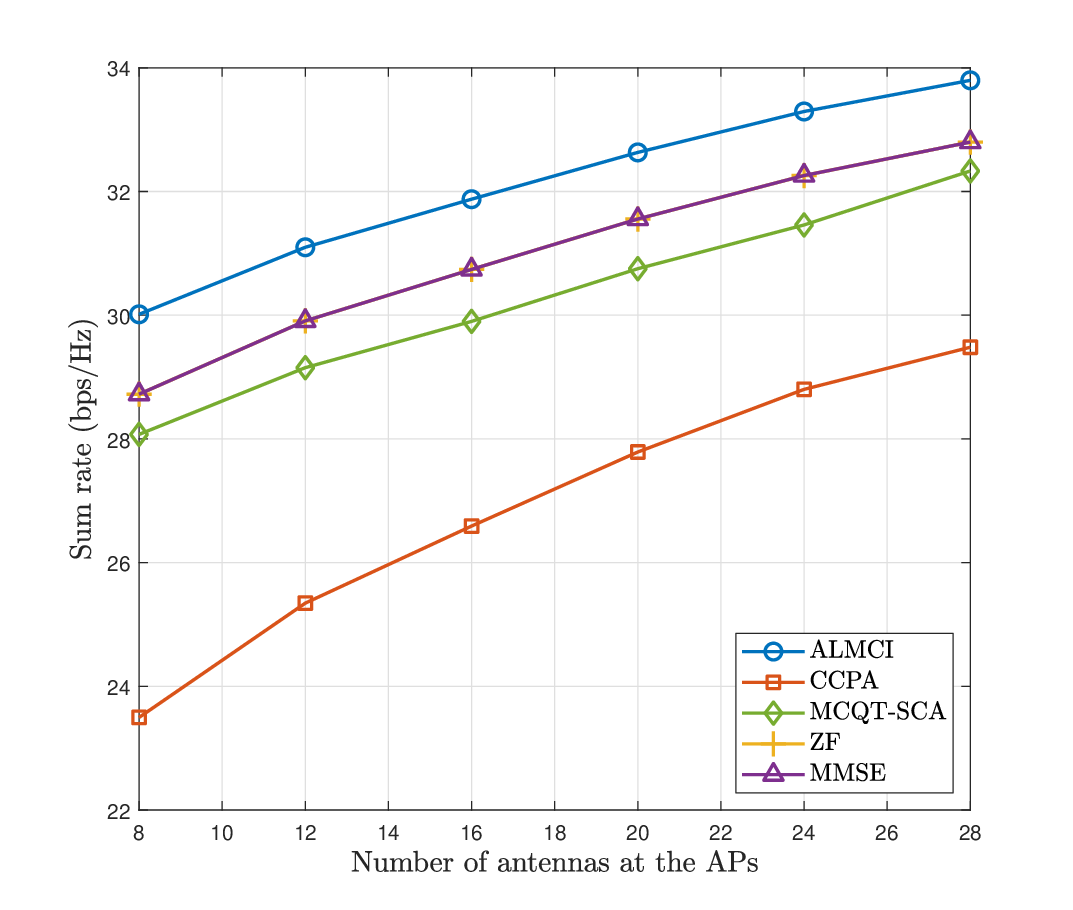}\vspace{-2mm} 
\caption{Sum rate versus the number of AP antennas, $L$. }
\vspace{-1mm}
\label{fig_Sumrate_antennas}
\end{figure}

Fig. \ref{fig_Sumrate_antennas} compares the achieved sum rates as a function of the number of AP antennas with $p_{\rm{max}}=\qty{30}{\dB m}$. While all the algorithms can leverage the spatial multiplexing gains,    ALMCI consistently outperforms the others, showing substantial gains as $L$ increases. Specifically, at $L=\num{12}$, it secures \qty{22.7}{\percent} and \qty{6.7}{\percent} higher sum rates than  CCPA and MCQT-SCA,  and a \qty{4.0}{\percent} gain over ZF and MMSE. Note that ZF and MMSE have virtually identical performance as the system operates in a high transmit power regime with a low noise variance, i.e., $\sigma^2 = \qty{-80}{\dB m}$.
Conversely, CCPA and MCQT-SCA perform poorer due to approximations and rank relaxations.

\begin{figure}[!t]\centering\vspace{-3mm}
\includegraphics[width=0.45\textwidth]{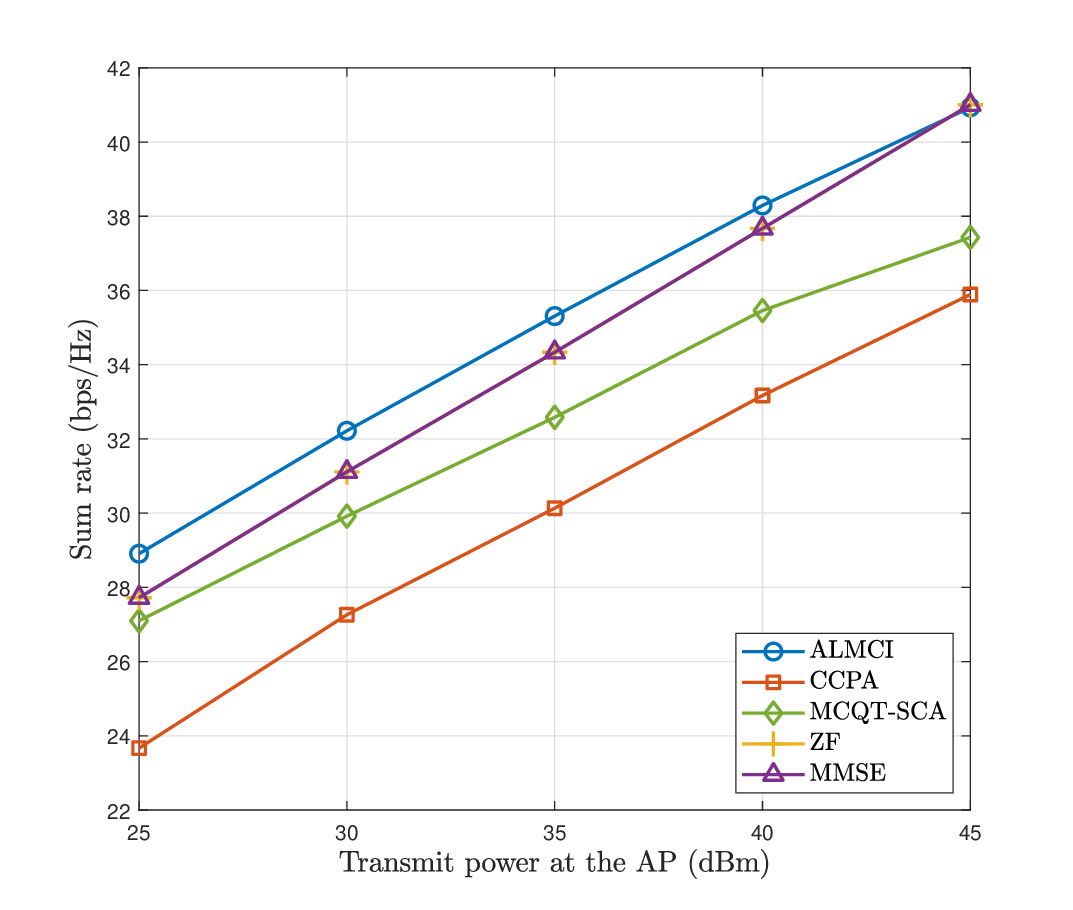}\vspace{-2mm} 
\caption{Sum rate versus the AP transmit power, $p_{\rm{max}}$. }\vspace{-1mm}
\label{fig_Sumrate_power}
\end{figure}

Fig.~\ref{fig_Sumrate_power} depicts the relationship between the sum rate and the AP transmit power, $p_{\rm{max}}$.  Once again, ZF and MMSE  perform identically across the evaluated power range. ALMCI  consistently achieves the highest sum rate. For instance, at $p_{\rm{max}}=$\qty{30}{\dB m}, it surpasses the CCPA, MCQT-SCA and ZF/MMSE by \qty{18.2}{\percent}, \qty{7.7}{\percent}, and \qty{3.6}{\percent}, respectively. All algorithms increase the sum rate with rising transmit power. Unlike ZF and MMSE, ALMCI considers sensing targets, contributing to higher power efficiency and greater rate gains per power unit. Meanwhile, CCPA and MCQT-SCA are less efficient in converting increased power to rate gains than others.  

\begin{figure}[!t]\centering\vspace{-4mm}
\includegraphics[width=0.45\textwidth]{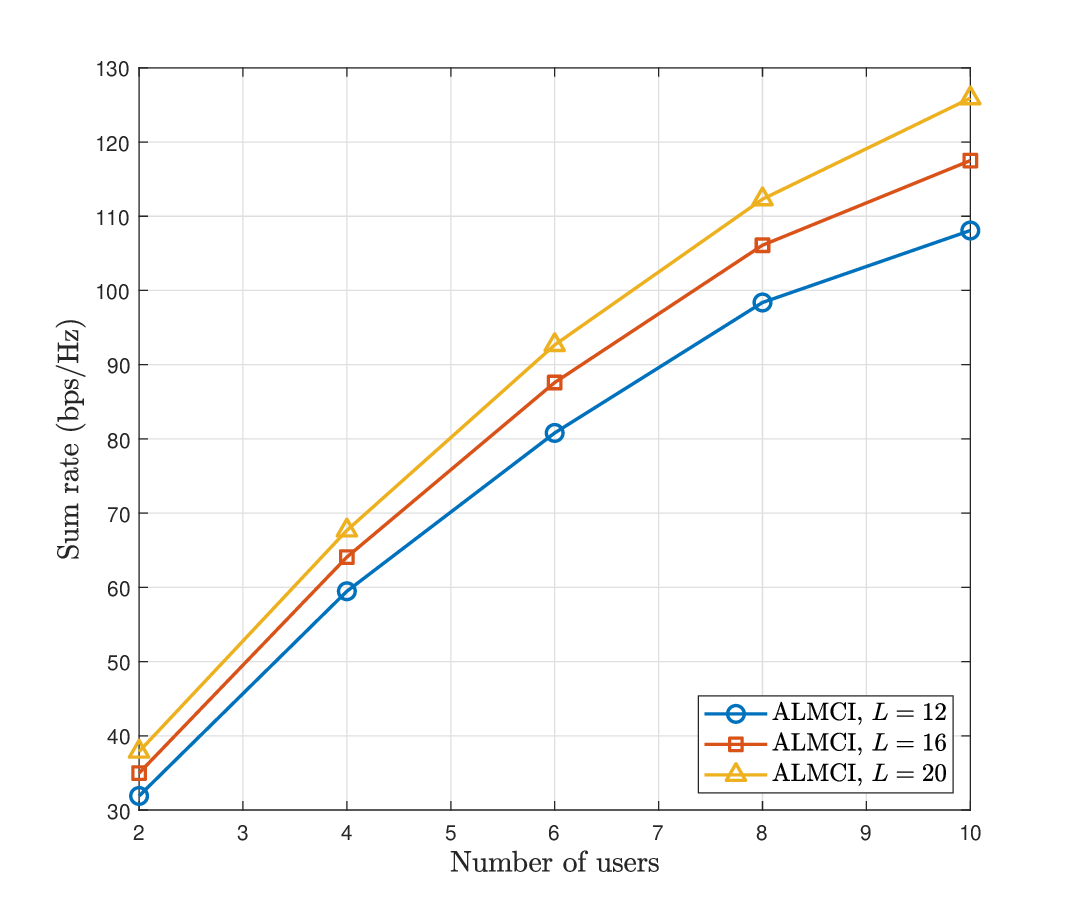}\vspace{-2mm} 
\caption{Sum rate versus the number of users, $K$. }\vspace{-3mm}
\label{fig_Sumrate_NumUser}
\end{figure}

Additionally, Fig.~\ref{fig_Sumrate_NumUser} shows the sum rate as a function of the number of users for ALMCI  with different values of $L$. Note that this figure includes only ALMCI as it has a lower time and computational complexity with any number of users (Fig.~\ref{fig_time_combine}). On the other hand, the algorithm execution times of CCPA and MCQT-SCA increase with the number of users. For example, generating a typical simulation figure with many users can take about a week, with time and computational complexity increasing exponentially with the number of users/APs/antennas.

In ALMCI, for all values of $L$, the sum rate increases as the number of users increases, illustrating the system's capacity to handle additional users efficiently. In addition, for a given number of users, a higher value of $L$ results in a higher sum rate. This indicates that the number of antennas can enhance the system's ability to effectively manage and distribute data across multiple users.

\section{Conclusion}
The concept of CF-ISAC is still in its early stages, with the literature being quite nascent. As a result, current CF-ISAC beamforming algorithms are suboptimal and may not scale well due to their failure to leverage the geometric characteristics of the problem. This paper addresses these limitations by developing a novel beamforming algorithm designed to maximize the communication sum rate while meeting sensing gain targets and AP power constraints. The key ideas are restricting the search to a complex oblique manifold and augmenting the objective function with a penalty term. These ensure that the search is on a smooth manifold and transmit power violations (8b) are handled.   The proposed ALMCI algorithm enhances beamforming precision and improves both the sum rate and spatial resolution. Extensive simulation demonstrates its superior performance compared to existing state-of-the-art methods.

Specifically, it is evaluated against  CCPA \cite{Mao2023, Demirhan2023, demirhan2024cellfree} and MCQT-CSA \cite{liu2024cooperative} methods. It notably slashed computational/time complexity and memory usage while boosting both sensing gain and sum rate. These improvements result in lower power consumption and hardware costs, particularly in large-scale networks, i.e., the reduced computational load on network hardware reduces energy consumption and cooling costs, extending hardware lifespan, improving resource utilization, and minimizing maintenance.  They also enable fast response to dynamic environments, allowing the network to support more concurrent users/targets \cite{Mai2022}.
Conversely, by leveraging problem geometry and Lagrangian penalty for constraint management, the ALMCI algorithm lays a solid foundation for efficiently tackling intricate optimization challenges, rendering it ideal for burgeoning large-scale networks.


As CF-ISAC research is still in its nascent stages, with limited literature \cite{Mao2023, Demirhan2023, demirhan2024cellfree, liu2024cooperative}, our approach opens up multiple research avenues, expanding the state-of-the-art. These include investigating the impact of imperfect CSI, limited backhaul capacity, user/target clustering, AP selection methods, and optimal pilot sequence design for concurrent channel estimation and target detection.

\bibliographystyle{IEEEtran}
\bibliography{IEEEabrv,ref}

\end{document}

%% file: SystemModel.eps_tex
\begingroup%
  \makeatletter%
  \providecommand\color[2][]{%
    \errmessage{(Inkscape) Color is used for the text in Inkscape, but the package 'color.sty' is not loaded}%
    \renewcommand\color[2][]{}%
  }%
  \providecommand\transparent[1]{%
    \errmessage{(Inkscape) Transparency is used (non-zero) for the text in Inkscape, but the package 'transparent.sty' is not loaded}%
    \renewcommand\transparent[1]{}%
  }%
  \providecommand\rotatebox[2]{#2}%
  \newcommand*\fsize{\dimexpr\f@size pt\relax}%
  \newcommand*\lineheight[1]{\fontsize{\fsize}{#1\fsize}\selectfont}%
  \ifx\svgwidth\undefined%
    \setlength{\unitlength}{761.17773438bp}%
    \ifx\svgscale\undefined%
      \relax%
    \else%
      \setlength{\unitlength}{\unitlength * \real{\svgscale}}%
    \fi%
  \else%
    \setlength{\unitlength}{\svgwidth}%
  \fi%
  \global\let\svgwidth\undefined%
  \global\let\svgscale\undefined%
  \makeatother%
  \begin{picture}(1,0.44433979)%
    \lineheight{1}%
    \setlength\tabcolsep{0pt}%
    \put(0,0){\includegraphics[width=\unitlength]{SystemModel.eps}}%
    \put(0.91850366,0.17554605){\color[rgb]{0,0,0}\makebox(0,0)[lt]{\lineheight{1.25}\smash{\begin{tabular}[t]{l}User-$k$\end{tabular}}}}%
    \put(0.15669089,0.32341564){\color[rgb]{0,0,0}\makebox(0,0)[lt]{\lineheight{1.25}\smash{\begin{tabular}[t]{l}User-$K$\end{tabular}}}}%
    \put(0.71907631,0.28250741){\color[rgb]{0,0,0}\makebox(0,0)[lt]{\lineheight{1.25}\smash{\begin{tabular}[t]{l}$\mathbf{h}_{mk}$\end{tabular}}}}%
    \put(0.67186781,0.18958387){\color[rgb]{0,0,0}\makebox(0,0)[lt]{\lineheight{1.25}\smash{\begin{tabular}[t]{l}$\mathbf{a}(\theta_{mn})$\end{tabular}}}}%
    \put(0.79743038,0.02230599){\color[rgb]{0,0,0}\makebox(0,0)[lt]{\lineheight{1.25}\smash{\begin{tabular}[t]{l}Target-$n$\end{tabular}}}}%
    \put(0.78682046,0.37309781){\color[rgb]{0,0,0}\makebox(0,0)[lt]{\lineheight{1.25}\smash{\begin{tabular}[t]{l}Target-$N$\end{tabular}}}}%
    \put(0.47475922,0.29984975){\color[rgb]{0,0,0}\makebox(0,0)[lt]{\lineheight{1.25}\smash{\begin{tabular}[t]{l}CPU\end{tabular}}}}%
    \put(0.5464112,0.02415145){\color[rgb]{0,0,0}\makebox(0,0)[lt]{\lineheight{1.25}\smash{\begin{tabular}[t]{l}AP-$m$\end{tabular}}}}%
    \put(0.23798121,0.00989636){\color[rgb]{0,0,0}\makebox(0,0)[lt]{\lineheight{1.25}\smash{\begin{tabular}[t]{l}AP-$1$\end{tabular}}}}%
    \put(0.01475118,0.15032192){\color[rgb]{0,0,0}\makebox(0,0)[lt]{\lineheight{1.25}\smash{\begin{tabular}[t]{l}AP-$M$\end{tabular}}}}%
    \put(0.04666361,0.01035841){\color[rgb]{0,0,0}\makebox(0,0)[lt]{\lineheight{1.25}\smash{\begin{tabular}[t]{l}Target-$1$\end{tabular}}}}%
    \put(0.38240589,0.02468931){\color[rgb]{0,0,0}\makebox(0,0)[lt]{\lineheight{1.25}\smash{\begin{tabular}[t]{l}User-$1$\end{tabular}}}}%
  \end{picture}%
\endgroup%